\documentclass[aps,prr,twocolumn,showpacs,final,floatfix,longbibliography]{revtex4-2}
\usepackage[pdftex]{graphicx}  % needed for figures
\usepackage[update]{epstopdf}
\usepackage{amssymb}
\usepackage{amsmath}
\usepackage{color}
\usepackage{textcomp}
\usepackage[utf8]{inputenc}

\def\d{\mathrm{d}\hspace*{-0.1ex}}
\graphicspath {{./figures/}}
\makeatletter
\def\input@path{{./figures/}}
\makeatother

\begin{document}

\title{Critical behavior of the three-state random-field Potts model in three dimensions}

\author{Manoj Kumar$^{1}$, Varsha Banerjee$^{2}$, Sanjay Puri$^{3}$, and Martin Weigel$^{1}$}
\affiliation{$^{1}$Institut für Physik, Technische Universität Chemnitz, 09107 Chemnitz, Germany.\\$^{2}$Department of Physics, Indian Institute of Technology, Hauz Khas, New Delhi -- 110016, India.\\$^{3}$School of Physical Sciences, Jawaharlal Nehru University, New Delhi -- 110067, India.}
% \affiliation{Department of Physics, Indian Institute of Technology, Hauz Khas, New Delhi -- 110016, India}

% \date{\today}

\begin{abstract}
Enormous advances have been made in the past 20 years in our understanding of the random-field Ising model, and there is now consensus on many aspects of its behavior at least in thermal equilibrium. In contrast, little is known about its generalization to the random-field Potts model which has wide-ranging applications. Here we start filling this gap with an investigation of the three-state random-field Potts model in three dimensions. Building on the success of ground-state calculations for the Ising system, we use a recently developed approximate scheme based on {\it graph-cut methods} to study the properties of the zero-temperature random fixed point of the system that determines the zero and non-zero temperature transition behavior. We find compelling evidence for a continuous phase transition. Implementing an extensive finite-size scaling (FSS) analysis, we determine the critical exponents and compare them to those of the random-field Ising model.
\end{abstract}

%\pacs{75.50.Lk, 64.60.F-, 02.60.Pn}
\maketitle
% systems on hypercubic lattice. 
% \section{Introduction}
% \label{s1}
{\it Introduction:}
Understanding the effect of quenched disorder on phase transitions is crucial for many experiments, such as magnetic systems with impurities, and technological application areas, such as quantum computers \cite{boixo:14}. At the same time, past progress in this direction has profoundly shaped the theory of equilibrium and non-equilibrium statistical mechanics, and theoretical concepts such as replica-symmetry breaking and the cavity method have found applications even in seemingly distant fields such as gene regulation \cite{decelle:11}, neural networks \cite{nishimori:book}, and the modelling of bird flocks \cite{bialek:12}.

Most of the recent focus has been on spin glasses, where competing and random interactions lead to a merely short-range ordered state, as well as on random-field systems \cite{young:book}. The latter are at the heart of such diverse problems as the behavior of the quantum magnet LiHo$_x$Y$_{1-x}$F$_4$ \cite{silevitch:07} and the random first-order transition scenario in structural glasses \cite{xia:00,guiselin:20}. For such problems, destruction of order is complete even for weak fields in $d=2$ dimensions \cite{imbrie:84,aizenmann:89a}, where ferromagnetic domains break up on length scales that vanish with growing strength of the random fields \cite{binder:83}. For continuous O($n$) spins, the lower critical dimension is even elevated to $d_\ell = 4$. The random-field Ising model (RFIM), on the other hand, orders at non-zero temperatures already for $d\ge3$, and the transition is of second order, at least for continuous field distributions \cite{fytas:13,fytas:16,kumar:17}. Hence the proposal of {\em dimensional reduction\/} \cite{aharony:76} suggested by field theory, where the RFIM in $d$ dimensions would be in the universality class of the $d-2$-dimensional ferromagnet, does not apply in low dimensions, but is only recovered for $d \ge d_c \approx 5$ \cite{fytas:17,tissier:11,kaviraj:21}.

Much less understood is the case of discrete spins with more than two states, i.e., the {\em random-field Potts model\/} (RFPM) \cite{nishimori,shapir84}. The Potts model has a plethora of applications ranging from finite-temperature QCD~\cite{alford:01}, over mixed antiferromagnets~\cite{domany:82}, orientational glasses~\cite{binder:92}, to soap froths~\cite{graner:92}. As disorder is inescapable, the RFPM is of even greater relevance for their study. Additionally, it is of profound theoretical interest since in the pure Potts model \cite{wu} the transition order can be tuned by changing the number of states $q$, such that there is a line $q_c^\mathrm{pure}(d)$ of tricritical points with $q_c^\mathrm{pure}(2)=4$ \cite{duminil:15b,duminil:16} and $q_c^\mathrm{pure}(3) \approx 2.35$  \cite{hartmann:05}. Since disorder tends to soften first-order transitions \cite{aizenmann:89a,cardy:99a}, one expects  a shift of the line $q_c^\mathrm{pure}(d)$ to $q_c^\mathrm{RF}(d)$. Just as for the  RFIM, dimensional reduction is not likely to hold in low dimensions \cite{shapir84}. Instead, one expects \cite{binder:83,imbrie:84} $d_\ell = 2$, and hence absence of long-range order in 2d --- a scenario that we recently confirmed numerically \cite{kumar2018approximate}. A plausible behavior of $q_c^\mathrm{RF}(d)$ is then $q_c^\mathrm{RF}(d\to2)\to\infty$ and $q_c^\mathrm{RF} = 2$ for $d \ge 6$ \cite{shapir84,ebjpcm}. Even once $q_c(d)$ is known, however, one needs to ask whether for all strengths $\Delta$ of random fields the first-order transition for $d_c^\mathrm{pure}(q) \le d \le d_c^\mathrm{RF}$ will be softened. This would be the case for $d=2$ \cite{aizenmann:89a}, but there is no FM order there. In $d > 2$, one might expect a line of tricritical points (or even two lines \cite{eb_epl}) to appear in the $(\Delta,T)$ plane, where the transition changes from first to second order \cite{goldschmidt:86,ebzpb}.

Very little is known about the details of this rich phase diagram (but see Ref.~\cite{turkoglu:21}). The purpose of this Letter is to address such issues for the  physically most relevant system of the $q=3$, $d=3$ RFPM, which in experiments has been used to describe trigonal-to-tetragonal structural transitions in SrTiO$_3$~\cite{aharony} when stressed along [111], and the mixed antiferromagnet Fe$_{1-x}$Co$_x$Cl$_2$ \cite{mukamel1981phase}. Such experimental systems show continuous transitions, but the theoretical situation is unclear as for this case $q_c^{RF} \ge q_c^\mathrm{pure} \approx 2.35$. Some early simulational work \cite{reed} found first-order transitions for all considered field strengths, but later studies claimed a continuous transition for intermediate fields combined with first-order behavior for small and large fields \cite{ebzpb,eb_epl}.  This scenario  agreed with the prediction of Ref.~\cite{shapir84} but contradicted Refs.~\cite{gx,*goldschmidt:86}, who performed a $1/q$-expansion of the $q$-state RFPM in 3D and found a first-order transition for $q\geq 3$, irrespective of the field strength. The question of a softening of the discontinuous transition hence has remained undecided.

Due to frustration and the ensuing slow relaxation, Monte Carlo methods are not very efficient in the presence of random fields. For the RFIM, much of the recent progress in understanding is due to the availability of efficient combinatorial optimization methods that allow to find exact ground states (GSs) in polynomial time \cite{dauriac:85,middleton02,stevenson:11,fytas:13}. Since the relevant renormalization-group fixed point is located at temperature $T=0$, such ground-state calculations are also relevant for the finite-temperature transitions. Unfortunately, the same methods do not extend to the RFPM, since  the ground-state problem is NP hard for $q > 2$ \cite{dauriac:85,bvz}. As we have recently shown, however, combinatorial graph-cut methods \cite{kolmogorov2004energy,kolmogorov:04} can still be used in combination with embedding techniques to efficiently compute high-quality approximate GSs \cite{kumar2018approximate,extrapolate}. To further improve the accuracy, we run the GS method for $n$ different random initial spin configurations and extrapolate the thermodynamic quantities in the limit $n \to \infty$. The extrapolated results enable us to uncover a clear-cut picture of the phase transition. 

%\begin{table}[tb]
%\centering
%\begin{tabular}{| c | c | c| c | c | c|c | c | c|c|}
 %\hline 
  %$L$&16& 20 &24&32&40&48&64&80&96\\ \hline
%$N_{\rm samp}/10^3$& 50&45&40&30&25&20&10&8&5\\ \hline
%$H(L)/10^{-2 }$ &8&5.72&4.355&2.83&2.024&1.54&1&0.72&0.544\\ \hline
% \end{tabular}
  %  \caption{Number of disorder samples ($N_{\rm samp}$) used for each system size $L$.
   % The last row shows the size of the uniform magnetic field $H(L)$ used to break the symmetry for determining the connected susceptibility $\chi$ ({\tt MW: do we need this?})
  % }
%\label{parameter}
%\end{table}

{\it Model and Methodology:}
We consider the $q$-state RFPM with Hamiltonian~\cite{shapir84}
\begin{equation}
\label{hamilt}
\mathcal{H}=-J\sum_{\left<ij\right>}\delta_{s_i,s_j}-\sum_i\sum_{\alpha=0}^{q-1}h_{i}^{\alpha}\delta_{s_i,\alpha},
\end{equation}
where $\delta_{x,y}$ is the Kronecker delta function, $s_i \in \{0,1,\ldots,q-1\}$, and $\{h_{i}^{\alpha}\}$ are uncorrelated, quenched random-field variables extracted
from a standard normal distribution, i.e., $P(h_i^\alpha) = (2\pi \Delta^2)^{-1/2} \exp [-(h_i^\alpha)^2/(2\Delta^2)]$;
$\Delta$ denotes the disorder strength. For $q=2$, Eq.~(\ref{hamilt}) maps to the RFIM at coupling $J/2$ and field strength $\Delta/\sqrt{2}$ \cite{kumar2018approximate}. Note that different couplings of random fields to the spins are possible as well as different coupling distributions \cite{goldschmidt:86,ebjpcm,turkoglu:21} but such variations are left for future work.

%%%%%%%%%%%%%%%%%%%%%%%%%%%%%%%%%%%%%%%%%%%%%%%%%%%%%%%%%%%
\begin{figure}[tb]
\begin{center}
\includegraphics[width=0.98\columnwidth]{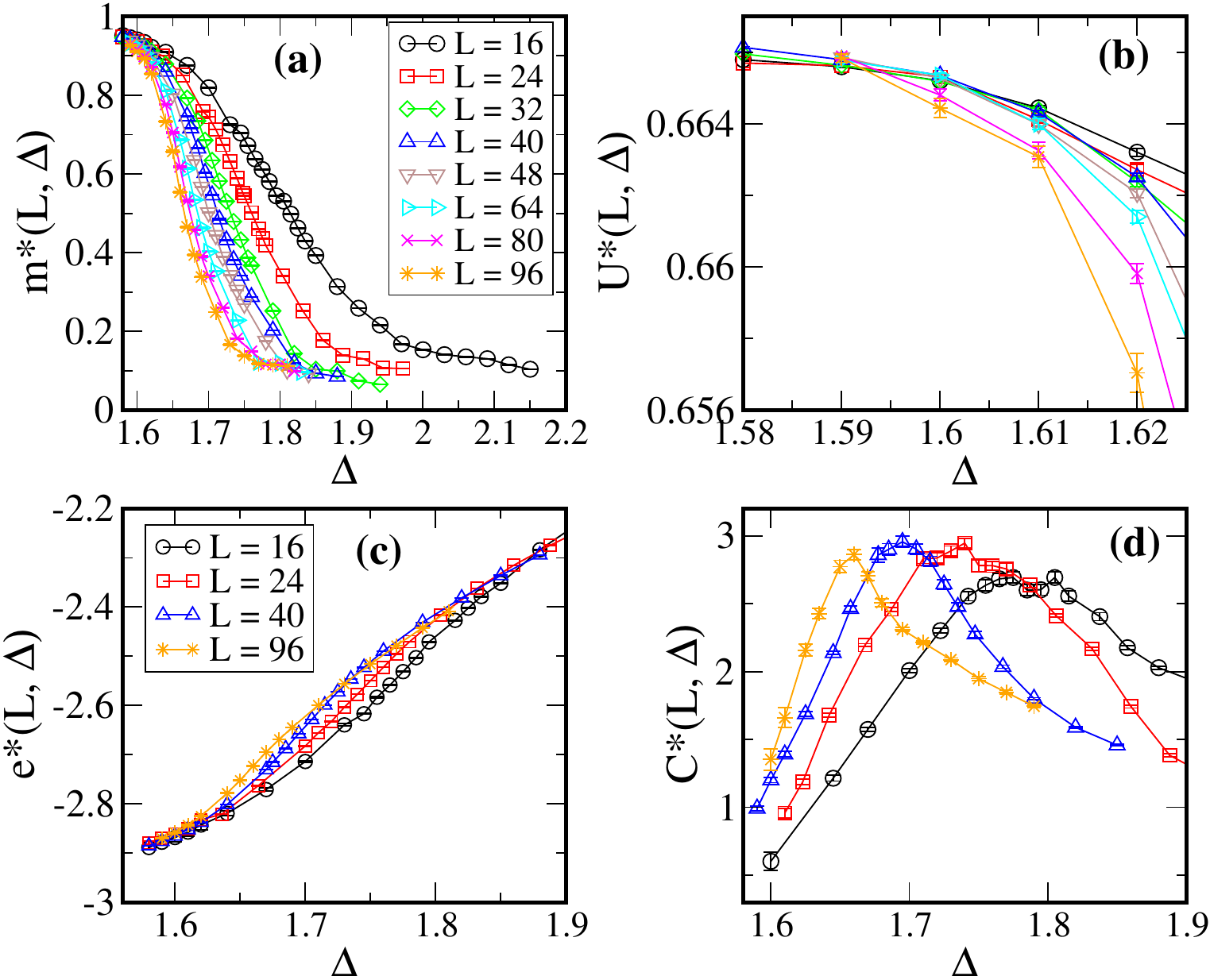}
\caption{Extrapolated estimates of the magnetization $m^*$, the Binder cumulant $U^*$, the bond energy $e^*$ as well as the specific heat  $C^*$ as a function of $\Delta$ for various system sizes $L$.
%Error bars are shown, which for $m^*$ and $e^*$ are much smaller than the symbol sizes.
}
\label{O_star}
\end{center}
\end{figure}
%%%%%%%%%%%%%%%%%%%%%%%%%%%%%%%%%%%%%%%%%%%%%%%%%%%%%%%%%%%%%%

We perform ground-state calculations for the $q=3$ RFPM on simple-cubic lattices of edge length $L$ with periodic boundary conditions.
%The number of disorder samples for each system size $L$ in the range $16\le L\le 96$ is indicated in Table~\ref{parameter}.
The number of disorder samples ranges from $N_\mathrm{samp} = 50\,000$ for $L=16$ to $N_\mathrm{samp} = 5\,000$ for $L=96$.
Approximate GSs are obtained using the algorithm described in Ref.~\cite{kumar2018approximate} that is based on an embedding of Ising spins into the Potts variables in the spirit of the $\alpha$-expansion method of Ref.~\cite{bvz}. For each disorder sample, we run our algorithm for $n$ different initial spin configurations and pick the run(s) resulting in the lowest energy as ground-state estimate. The success probability of the resulting approach increases exponentially with $n$, such that the method becomes exact for $n\to\infty$ \cite{extrapolate}. For each sample we determine the order parameter \cite{wu}
 \begin{equation}\label{psi1}
 m (L,\Delta, n)= \frac{q\rho-1}{q-1},~\text{where}~ \rho=\frac{1}{L^3} \max_{\alpha} \sum_{i}\delta_{s_i,\alpha}.
\end{equation}
Here, $\rho$ denotes the density of spins in the majority orientation. Also, we measure the bond energy per spin $e_J(L,\Delta, n)= - \sum_{\langle ij \rangle} \delta_{s_i,s_j}/L^3$~\cite{hartmann:01b}.
%\begin{equation}
% e_J(L,\Delta, n)= - \frac{1}{N}\sum_{\langle ij \rangle} \delta_{s_i,s_j}. 
%\end{equation}
After performing the disorder average, $[\cdot]_{\rm av}$, we then deduce further quantities such as the Binder cumulant associated to $m$, $U_{4} (L,\Delta, n) = 1 - [m^{4}]_{\rm av}/3[m^{2}]_{\rm av}^{2}$.
The employed definitions of the specific heat and magnetic susceptibility will be discussed below. All statistical errors were estimated using the jackknife method \cite{efron82,miller74,weigel:10}. Estimates for $\Delta_c$ and the critical exponents are then extracted from the scaling of observables at sequences of pseudocritical points as well as from scaling collapses~\cite{melchert09}. 

\begin{table}[tb]
\centering
    \caption{Estimates of $\Delta_c$, $\nu$, and $\beta/\nu$ according to Eq.~\eqref{eq:mscaling} as well as $\bar{\gamma}/\nu$ according to Eq.~\eqref{eq:chidis-scaling} extracted from scaling collapses of the data for different $n$ as well as the extrapolated data for $n\to\infty$ ($L_\mathrm{min} = 24$). $S_1$ and $S_2$ are the qualities of the collapses according to \eqref{eq:mscaling} and \eqref{eq:chidis-scaling}, respectively ($S\approx 1$ for perfect collapses).}
\label{mag_exp}
\begin{ruledtabular}
\begin{tabular}{ c  c  c  c c c c }
 
 $n$&$\Delta_c$&$1/\nu$&$\beta/\nu$&$\bar{\gamma}/\nu$ &$S_1$&$S_2$\\ \hline 
 1&1.636(2)&0.837(9)&0.0460(9)&2.9084(14)&2.30&2.38 \\ 
 5&1.626(3)&0.812(6)&0.0403(8)&2.9220(15) &1.82&1.69 \\ 
 10&1.623(5)&0.828(15)&0.0387(7)&2.9230(15)&1.28&1.58 \\
 50&1.617(4)&0.797(4)&0.0340(8)&2.9323(16)&1.25&1.38 \\
100&1.616(1)&0.774(6)&0.0330(10)&2.9337(15)&1.20&1.36 \\ 
$\infty$&1.606(3)&0.723(4)&0.0306(23)&2.9402(30)&0.82&0.87 \\ 
 \end{tabular}
 \end{ruledtabular}
\end{table}

{\it Extrapolation and transition order:} To assess the quality of approximation, we first studied the behavior of each quantity as a function of $n$. We generally find a two-stage behavior, with an initial fast decay followed by a much slower large-$n$ convergence, which is well described by the sum of two power laws \cite{extrapolate},
\begin{equation}
\label{extrapolate_form}
O(L,\Delta, n)= a n^{-b}(1+c n^{-e})+O^*(L,\Delta),
\end{equation}
where $b < e$ is the {\it asymptotic}, slow exponent, $e$ describes the initial fast decay, and $O^*$ denotes the limiting value for $n\to \infty$. As was shown elsewhere, this form is quite generic and it holds, in particular, for a certain subset of samples for which exact GSs are known \cite{extrapolate}. For such exact samples of size $16^3$ we employed our algorithm for up to $n = 10^4$ runs and found that the residuals with respect to the exact results, i.e., $O(n)-O_{\rm ex}$ for any quantity scale as $a n^{-b}(1+c n^{-e})$, with $b\simeq 0.02$ and $e\simeq 0.5$. This behavior is seen to extend to the case where the exact results are not used or known \cite{extrapolate}. The value of $b$ is found to be very stable in this regard, such that we fix it for the subsequent fits of our main study reported here, for which $n \le 100$ \footnote{We  have explicitly verified that a slight variation in the exponent $b$ does not yield significantly different results in the extrapolations.}. We then perform joint fits of  the functional form \eqref{extrapolate_form} to $[m]_{\rm av}$, $U_4$, and $[e_J]_{\rm av}$ for a common value of the exponent $e$, yielding extrapolated estimates $m^*$, $U^*$, and $e^*$ for any fixed $(L, \Delta)$ (see Section S1 of Ref.~\cite{SM}). We prefix our analysis by a study of the energetic Binder cumulant, whose scaling clearly shows the behavior expected from a continuous transition, as already reported for $T > 0$ in Ref.~\cite{ebjpcm}; details are provided in Ref.~\cite{SM}, Section S2. In the following, we perform FSS of all quantities for finite $n$ as well as for $n\to\infty$, in order to determine the transition point and  obtain the critical exponents.

%We will see that the exponents for infinite $n$ precisely describe the critical behavior of the system when compared to those for finite $n$.     
% $Q$ determines the probability that the value of 
% \begin{equation}
%  \chi^2=\sum_{i}^{N}\left(\frac{y_i-g(x_i)}{\sigma_i}\right)^2,
% \end{equation}
% with $N$ data points $(x_i,y_i\pm \sigma_i)$ fitted to the function $g$, is worse than the current fit~\cite{hartmann:01a,press96}. If $Q\gtrsim 0.1$,  then the goodness-of-fit is believable. If $Q\gtrsim 0.001$, then the fit may be acceptable if the errors are nonnormal or have been moderately underestimated. If $Q < 0.001$, then the fit is not acceptable (see, numerical recipes in C, Ref.~\cite{press96}).
%   

\begin{figure}[tb]
\begin{center}
\includegraphics[width=0.98\columnwidth]{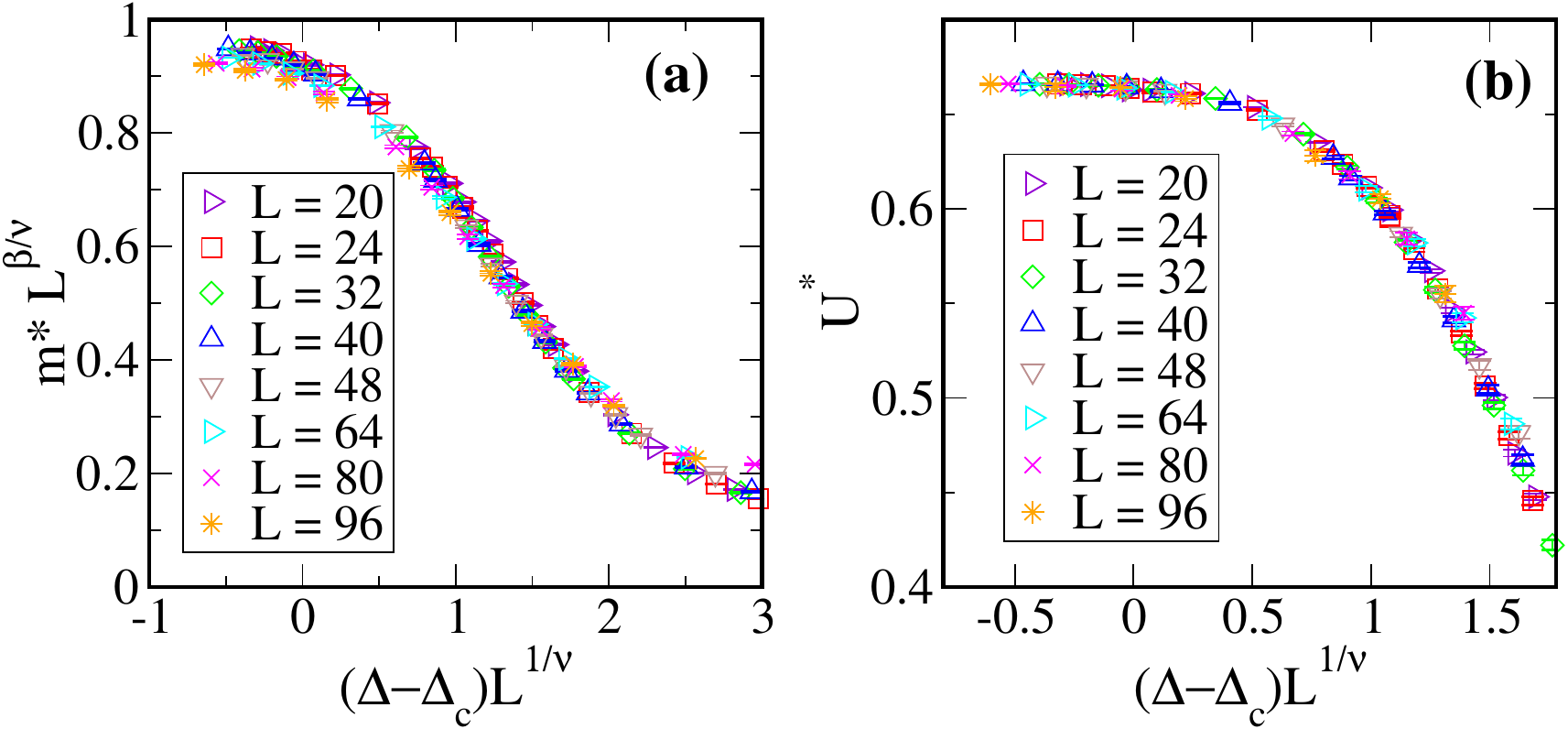}
\caption{(a) Scaling plot of $m^*(L,\Delta)L^{\beta/\nu}$ versus $(\Delta-\Delta_c)L^{1/\nu}$ with  $
\Delta_c=1.606$, $1/\nu=0.723$, $\beta/\nu=0.0306$. (b) Scaling collapse of $U^*$ vs. $(\Delta-\Delta_c)L^{1/\nu}$.}
\label{mag_cum_scale}
\end{center}
\end{figure}

{\it Magnetization and Binder cumulant:} In panel (a) of Fig.~\ref{O_star} we show the extrapolated  magnetization $m^*$ as a function of the disorder strength $\Delta$ for various lattice sizes $L$. The expected FSS form \cite{privman:privman}
\begin{equation}
 m^*(L, \Delta)= L^{-\beta/\nu}\widetilde{ \mathcal M}\left[(\Delta-\Delta_c)L^{1/\nu}\right],
  \label{eq:mscaling}
\end{equation}
implies that a plot of $m^*(L,\Delta) L^{\beta/\nu}$ against $x = (\Delta-\Delta_c)L^{1/\nu}$ should yield a collapse of data sets for small $|x|$ for appropriate values of $\Delta_c$, $\nu$, and $\beta/\nu$. This is shown in Fig.~\ref{mag_cum_scale}(a).
% The finite-size scaling is performed using the python program autoscale.py~\cite{melchert09} with the initial choices of parameters $\Delta_c=1.62$, $1/\nu=0.8$, $\beta/\nu=0.1$.
The collapse with the least mean-squared deviation $S$ from the master curve \cite{houdayer:04,melchert09} is obtained for $\Delta_c=1.606\pm 0.003, 1/\nu=0.723\pm 0.004$, and $\beta/\nu=0.0306\pm 0.0023$. We also performed FSS of the magnetization for finite $n=1$, $5$, $10$, $50$, and $100$. As is clear from Table~\ref{mag_exp}, there is a weak dependence of the estimates on $n$ with a smooth convergence to the exact limit $n\to \infty$.

Turning to the Binder cumulant, from Fig.~\ref{O_star}(b) we see a crossing of $U^*(L,\Delta)=U_4(L,\Delta, n\to \infty)$ in the range $\Delta_{c} = [1.59,1.61]$, predicting the location of the critical point, where $U_4$ becomes system-size independent \cite{binder81, binder:book1}. The FSS form of $U^*$ is \cite{privman:privman}
$
 U^*(L, \Delta)= \widetilde{ \mathcal U}\left[(\Delta-\Delta_c)L^{1/\nu}\right].
$
As is seen from the rescaled data in Fig.~\ref{mag_cum_scale}(b), a rather clean scaling collapse is achieved for $\Delta_c = 1.604(2)$ and $1/\nu=0.720(6)$, yielding an alternative, and consistent, set of estimates for $\Delta_c$ and $1/\nu$.

{\it Specific heat:} Due to the restriction to $T=0$ and the uniqueness of the ground state, it is not possible to define the specific heat from a temperature derivative or fluctuation-dissipation relation. Instead, a specific-heat like quantity is given by the derivative $C(\Delta)=\partial [e_J(\Delta)]_{\rm av}/\partial \Delta$ \cite{hartmann:01b,fytas:16b}. Numerically, we compute this using the standard three-point formula at the midpoint~\cite{numrec}.
%, i.e., if $\Delta_1$ and $\Delta_2$ are the two nearby values of $\Delta$, then 
%\begin{equation}\label{sp_heat}
 %C\left(\frac{\Delta_1+\Delta_2}{2}\right)=\frac{[e_J(\Delta_2)]_{\rm av}-[e_J(\Delta_1)]_{\rm %av}}{\Delta_2-\Delta_1}.
%\end{equation}
In Fig.~\ref{O_star}(d) we show the extrapolated $C^*(L,\Delta)$. As $L$ is increased, the peaks shift but only weakly vary in height, indicating a small specific-heat exponent $\alpha$. To determine the latter, we considered additional $\Delta$ values and used parabolic fits  $C(L, \Delta)=a_0(\Delta-\Delta_{\max,C})^2+C_{\max}$ to obtain the peak locations $\Delta_{\max,C}(L)$ and peak heights $C_{\max}(L)$. In a finite system, the {\it singular} part of $C(L,\Delta)$ scales as \cite{privman:privman}
$
 C_s(L,\Delta) = L^{\alpha/\nu}\widetilde{C}\left[(\Delta-\Delta_c)L^{1/\nu}\right].
$
If the maximum of $\widetilde{C}$ occurs for argument $a_1$, then the positions in $\Delta$ shift as
\begin{equation}
\label{deltac}
 \Delta_{\max,C}(L)\approx\Delta_c+a_1 L^{-1/\nu},
\end{equation}
and the maximum value of the singular part of the specific heat $C_{s,\mathrm{max}}(L)\sim L^{\alpha/\nu}$.

%%%%%%%%%%%%%%%%%%%%%%%%%%%%%%%%%%%%%%%%%%%%%%%%%%%%%%%
%\begin{figure}
%\begin{center}
%\includegraphics[width=0.98\columnwidth]{cmax_delta_c}
%\caption{ Plots of (a) $\Delta^{\rm ps}(L)$ where the specific-heat attains its maximum and (b) the maximum of specific heat $C^{\max}(L)$ as a function of $L$ for different $n$. The solid curves in (a) and (b) are the  best fit to Eqs.~\eqref{deltac} and \eqref{cmax_scale} with the values of $\Delta_c$ and exponents $1/\nu$, $\alpha/\nu$ and $\omega$  collected in Table~\ref{sp_heat_exp}. }
%\label{delta_c_sp_heat}
%\end{center}
%\end{figure}
%%%%%%%%%%%%%%%%%%%%%%%%%%%%%%%%%%%%%%%%%%%%%%%%%

\begin{figure}[tb]
\begin{center}
\includegraphics[width=\columnwidth]{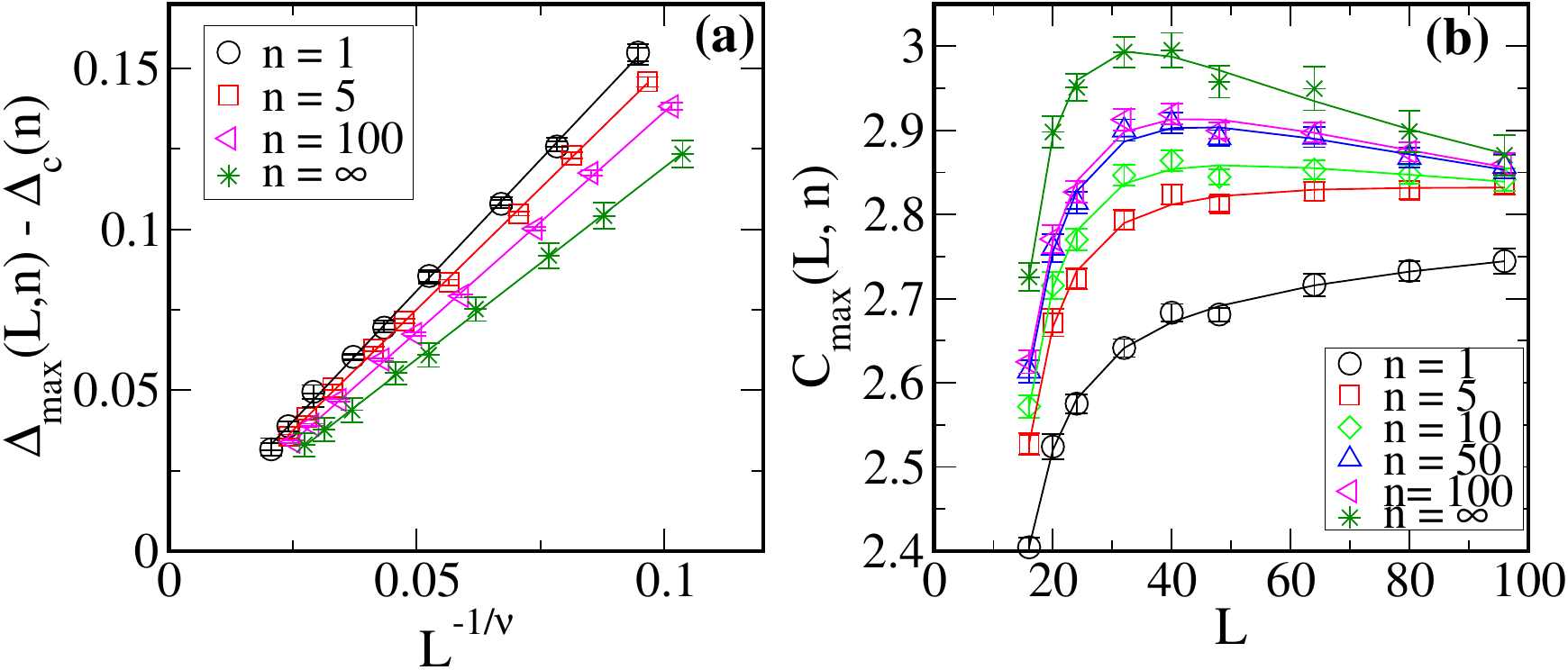}
\caption{(a) Shifts $\Delta_{\max,C}(L)-\Delta_c$ of the pseudo-critical fields of the specific heat against $L^{-1/\nu}$ for different $n$, where $\Delta_c$ and the exponent $1/\nu$ are determined from fits of the form \eqref{deltac}, see Table~\ref{sp_heat_exp}. For increased clarity, the data for different $n$ are slightly shifted relative to each other. (b) Scaling of the maxima $C_{s,\max}(L)$ as a function of $L$ for different $n$. The curves correspond to fits of the form \eqref{cmax_scale} to the data (cf.
Table~\ref{sp_heat_exp}).}
\label{delta_c_sp_heat}
\end{center}
\end{figure}
%%%%%%%%%%%%%%%%%%%%%%%%%%%%%%%%%%%%%%%%%%%%%%%%%
% According to Eq.~\eqref{deltac}, the pseudo-critical points $\Delta^{\rm ps}(L)$, determined from Fig.~\ref{O_star}(d), can be used to estimate the infinite-size critical disorder $\Delta_c$ (see Fig.~\ref{delta_c}).

Figure~\ref{delta_c_sp_heat}(a) shows our data for $\Delta_{\max,C}(L)$ together with fits of the form \eqref{deltac}, indicating clear consistency. The estimates of $\Delta_c$ and the exponent $1/\nu$ are collected in Table~\ref{sp_heat_exp}, where we also indicate the quality $Q_1$ of these fits \cite{numrec}. In Fig.~\ref{delta_c_sp_heat}(b) we present the corresponding peak heights $C_{\max}(L,n)$. This plot clearly suggests that $\alpha$ is either positive, but very small, or negative. As a consequence, scaling corrections are relevant and we hence considered the functional form
\begin{equation}\label{cmax_scale}
 C_{\max}(L)=C_0 + c_1 L^{\alpha/\nu} (1+c_2L^{-\omega}),
\end{equation}
where $\omega$ corresponds to the Wegner exponent and $C_0$ represents a non-singular background term. Since $\alpha \approx 0$ would result in a second additive constant in \eqref{cmax_scale}, we cannot reliably include all five parameters in the fit. We hence fix $C_0 = 0$, and the resulting 4-parameter fit yields excellent qualities $Q_2$ and the estimates of $\alpha/\nu$ and $\omega$ collected in Table~\ref{sp_heat_exp}. We thus conclude that $\alpha$ is very slightly negative or perhaps zero.

%\begin{table} 
%\centering
%\begin{tabular}{| c | c | c | c| c|c | c|}
 %\hline 
 %$n$&$\Delta_c$&$1/\nu$&$\alpha/\nu$&$\omega$&$Q_1$&$Q_2$ \\ \hline
 %1&1.645(9)&0.86(11)&$0.009(30)$&$1.74(97)$ &0.83&0.92 \\ \hline
% 5&1.629(4)&0.814(48)&$-0.004(21)$& $2.13(89)$& 0.34&0.91 \\ \hline
 %10&1.626(3)&0.835(38)&$-0.026(33)$&$1.65(83)$ &0.14&0.7\\ \hline
 %50&1.623(4)&0.812(43)&$-0.061(45)$&$1.39(72)$ &0.12&0.99 \\ \hline
%100&1.625(3)&0.827(43)&$-0.072(51)$&$1.3(7)$ &0.11&0.97\\\hline
%$\infty$&1.610(6)&0.80(7)&$-0.08(7)$&$1.76(93)$&0.91&0.87\\ \hline
 %\end{tabular}
    %\caption{Parameters of fits of the functional forms \eqref{deltac} and \eqref{cmax_scale} to the data for the %specific heat.}
%\label{sp_heat_exp}
%\end{table}

%%%%%%%%%%%%%%%%%%%%%%%%%%%%%%%%%%%%%%%%%%%%%%%%%%%%%%%%%%%%%%
\begin{table}[b] 
\centering
    \caption{Parameters of fits of the functional forms \eqref{deltac} and \eqref{cmax_scale} to the data for the specific heat. $Q_1$ and $Q_2$ refer to the quality-of-fit \cite{numrec}. 
    %({\tt MW: why do we show quality-of-fit here, but not in Table II?})
    }
\label{sp_heat_exp}
\begin{ruledtabular}
\begin{tabular}{ c  c  c  c  c c  c }
  $n$&$\Delta_c$&$1/\nu$&$\alpha/\nu$&$\omega$&$Q_1$&$Q_2$ \\ \hline
 1&1.644(6)&0.850(70)&$0.023(12)$&$2.67(87)$ &0.74&0.71 \\ 
 5&1.626(3)&0.774(32)&$-0.002(11)$& $2.62(68)$& 0.32&0.70 \\
 10&1.621(3)&0.767(25)&$-0.019(13)$&$2.39(61)$ &0.14&0.52\\ 
 50&1.620(2)&0.776(21)&$-0.046(20)$&$1.87(53)$ &0.12&0.50 \\
100&1.620(2)&0.780(21)&$-0.049(20)$&$1.86(52)$ &0.15&0.49\\
$\infty$&1.611(4)&0.733(28)&$-0.059(20)$&$2.52(73)$&0.14&0.93\\
 \end{tabular}
 \end{ruledtabular}
\end{table}

\begin{figure}[tb]
\begin{center}
\includegraphics[width=0.98\columnwidth]{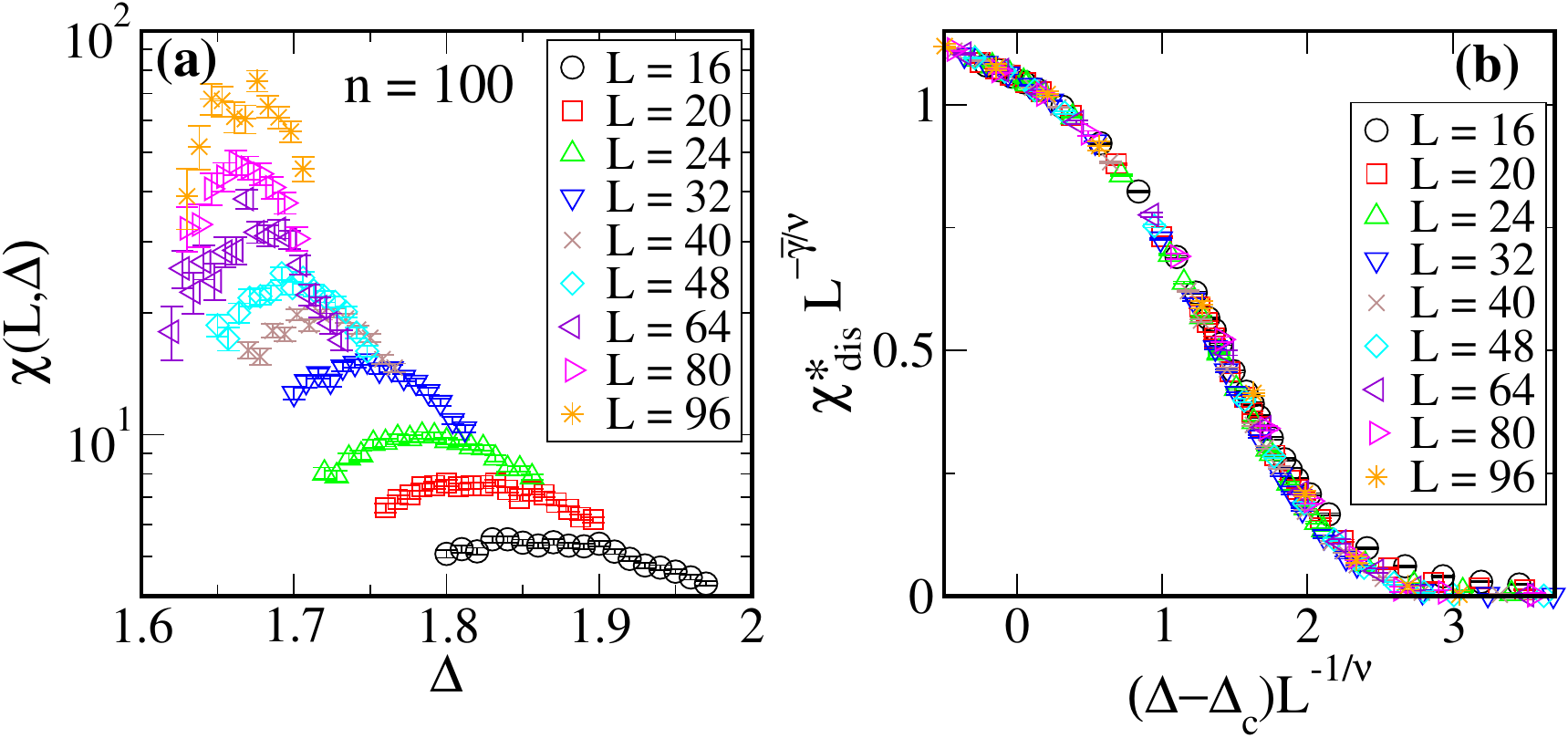}
\caption{(a) $\chi(L,\Delta,n=100)$ (on a log-linear scale) against $\Delta$ and for different $L$ (see key). (b) Data collapse of $\chi^*_{\rm dis} (L, \Delta)$ with the exponent values in Table~\ref{mag_exp}. }
\label{chi_rfpm}
\end{center}
\end{figure}

{\it Susceptibility:}
We now turn to the connected and disconnected magnetic susceptibilities. Since we operate at $T=0$, we cannot use the usual fluctuation-dissipation relation. As outlined in Section S3 of Ref.~\cite{SM}, we generalize arguments for the RFIM \cite{schwartz1985exact} to express the zero-field RFPM susceptibility,
\begin{equation}
  \left[ \frac{\partial \langle M^\mu\rangle}{\partial H^\mu}\right]_\mathrm{av} =
  \int\d\{\bar{h}_i^\alpha\} P(\{\bar{h}_i^\alpha\})  \frac{\partial \langle
    M^\mu\rangle_{\bar{h}^\alpha_i}}{\partial H^\mu},
  \label{eq:schwartz1}
\end{equation}
for Gaussian random fields as $ \chi^\mu = [ \langle m^\mu\rangle \sum_i h_i^\mu]_\mathrm{av}/\Delta^2$, where $m^\mu =   \sum_i \delta_{s_i,\mu}/L^3$ (see also \cite{fytas:16}). We apply a constant external field $H$  to the spin state 1  (i.e., $\mu=1$) to break the symmetry so that $\chi$ displays a peak. As is shown in S3, a minimal field strength $\propto L^{-3/2}$ is necessary to break the symmetry, and we choose $H(L=16) = 8\times 10^{-2}$. 
Due to the large sample-to-sample fluctuations in $\chi$ we do not observe a systematic variation of $\chi(L,\Delta,n)$ with $n$, such that instead of an extrapolation we focus on $n=100$.
Figure~\ref{chi_rfpm}(a) shows the behavior of $\chi(L,\Delta, n=100)$ as a function of $\Delta$ and for various $L$. To analyze its divergence, we fit a parabola near the peak and obtain the peak positions $\Delta_{{\max},\chi}(L)$ as well as the maxima of the susceptibility $\chi_{\max}(L)$. FSS predicts that
\begin{equation}
 \label{eq:susceptibility-scaling}
 \chi(L,\Delta) =L^{\gamma/\nu} \widetilde{\mathcal{\chi}}\left[(\Delta-\Delta_c)L^{1/\nu}\right], ~\chi_{\max}(L) \sim L^{\gamma/\nu}. 
\end{equation}

%\begin{figure}[tb]
%\begin{center}
%\includegraphics[width=0.98\columnwidth]{chi_rfpm_ss}
%\caption{(a) $\chi(L,\Delta,n=100)$ (on a log-linear scale) against $\Delta$ and for different $L$ &(see key). (b) Susceptibility maxima $\chi_{\max}(L)$ as a function $L$. The solid line is a %power-law fit $\chi_{\max}(L)=AL^{\gamma/\nu}$, yielding $\gamma/\nu = 1.36 \pm 0.01$ with %fit-quality $Q=0.034$. The inset shows $\Delta_{\max,\chi}(L)$ versus $L$. The solid curve %corresponds to a fit $\Delta_{\max,\chi}(L) =\Delta_c+a_1 L^{-1/\nu}$ with $\Delta_c=1.621(5)$, %$1/\nu=0.97(5)$ and fit-quality $Q=0.007$. }
%\label{chi_rfpm}
%\end{center}
%\end{figure}

A power-law fit of the form \eqref{eq:susceptibility-scaling} for $\chi_\mathrm{max}(L)$ yields $\gamma/\nu=1.36(1)$ ($Q=0.034$). Similarly, using the form  $\Delta_{\max,\chi}(L)=\Delta_c+a_1L^{-1/\nu}$ for the peak locations yields the estimates $\Delta_c=1.621(5)$ and $1/\nu=0.97(5)$ ($Q=0.007$). The corresponding data are shown in Fig.~S5(a) of Ref.~\cite{SM}. While the value of $\Delta_c$ agrees with that obtained from the specific heat (see Table~\ref{sp_heat_exp} for $n=100$), the estimate for $1/\nu$ is noticeably larger than the values from the magnetization and specific heat, indicative  of unresolved scaling corrections. To avoid this problem, we performed additional simulations without external field $H$ at {\em fixed\/} $\Delta \in [1.602,1.606]$, corresponding to the estimate $\Delta_c = 1.604(2)$ found above from $U^\ast(L,\Delta)$. Here, we find clean scaling, cf.\ Fig.~S5(b), and the fit parameters collected in Table S1, indicating little scaling corrections in $\chi$, and the dependence of $\gamma/\nu$ on $\Delta$ is only weak; we quote $\gamma/\nu=1.51(6)$  at our best estimate $\Delta_c=1.604$ for $L_\mathrm{min} = 16$. 

%\begin{table*}[tb]
%\centering
%\begin{tabular}{| c | c | c|c|c|c|c|c|c|c|c|}
 %\hline 
%$L_{\min}$ & \multicolumn{2}{|c|}{$\Delta_c=1.602$} & \multicolumn{2}{|c|}{$\Delta_c=1.603$} %&\multicolumn{2}{|c|}{$\Delta_c=1.604$} & \multicolumn{2}{|c|}{$\Delta_c=1.605$} %&\multicolumn{2}{|c|}{$\Delta_c=1.606$}\\
%\cline{2-11}
%&$\gamma/\nu$&$Q$&$\gamma/\nu$&$Q$&$\gamma/\nu$&$Q$&$\gamma/\nu$&$Q$&$\gamma/\nu$&$Q$\\ \hline
%16& 1.64(6)& 0.47 &1.54(6)&0.49&1.512(58)&0.68&1.59(6)&0.67&1.463(61)&0.67\\ \hline
%20&1.635(73)& 0.36 &1.512(74)&0.43&1.467(72)&0.71&1.583(74)&0.56&1.453(76)&0.56 \\ \hline
%24& 1.58(9)& 0.35 &1.60(9)& 0.57&1.493(91)& 0.61&1.595(92)&0.43&1.380(96)&0.66 \\ \hline
%32& 1.56(13)&0.24&1.69(13)& 0.55&1.62(13)& 0.78&1.55(13)& 0.33&1.44(14)&0.58\\ \hline
%\end{tabular}
    %\caption{The exponent $\gamma/\nu$ from fits of the form $\chi(L,\Delta_c) \sim L^{\gamma/\nu}$ for different %estimates $\Delta_c$ of the critical field strength as a function of the cut-off $L_{\min}$.}
 %\label{gamma_exp}
%\end{table*}

The additional length scale introduced by the disorder fluctuations results in a different scaling behavior of the static, disconnected susceptibility $\chi_{\rm dis} = L^d[m^2]_{\rm av}$ as compared to  the thermal, connected susceptibility $\chi$ \cite{fisher1986scaling},
\begin{equation}
    \chi_{\rm dis}(L,\Delta) =L^{\bar\gamma/\nu} \widetilde{\mathcal \chi}_{\rm dis}[(\Delta-\Delta_c)L^{1/\nu}].
    \label{eq:chidis-scaling}
\end{equation}
We use this relation to perform a scaling collapse for the extrapolated  $\chi^*_{\rm dis}$. As is shown in Fig.~\ref{chi_rfpm}(b), this leads to an excellent scaling result; the exponents $\bar \gamma/\nu$, including those for finite $n$, are provided in Table~\ref{mag_exp}, with $\bar{\gamma}/\nu = 2.9402(30)$ for $n=\infty$. The so-called two-exponent scaling scenario predicts $\bar \gamma = 2\gamma$ \cite{schwartz1985random,schwartz1991missing}. From our data we find the marginal result $(2\gamma-\bar{\gamma})/\nu = 0.08(6)$. We can also investigate the validity of the Rushbrooke equality $\alpha+2\beta+\gamma=2$ and the {\it modified} hyperscaling relation $2-\alpha=\nu (d-\theta)$ \cite{schwartz1991breakdown,grinstein1976ferromagnetic,fisher1986scaling}, where $\theta=2-\bar \eta+\eta= (\bar \gamma-\gamma)/\nu$. It can be inspected from Tables~\ref{mag_exp} and \ref{sp_heat_exp} that both relations are well satisfied (within error bars) by the results for infinite $n$. 

\begin{table}[tb!]
    \centering
        \caption{Critical exponents of the 3D $q=3$ RFPM as compared those of the RFIM \cite{fytas:17}.}
    \label{tab:exponents}
    \begin{ruledtabular}
    \begin{tabular}{crr}
         &  RFIM & $q=3$ RFPM\\ \hline
         $\nu$ & 1.38(10) &  1.383(8) \\
         $\alpha$ & -0.16(35) & -0.082(28) \\
         $\beta$ & 0.019(4) & 0.0423(32) \\
         $\gamma$ & 2.05(15) & 2.089(84) \\
         $\eta$ & 0.5139(9) & 0.49(6) \\
         $\bar{\eta}$ & 1.028(2) & 1.060(3) \\
         $\theta$ & 1.487(1) & 1.43(6) \\
         $\alpha+2\beta+\gamma$ & 2.00(31) & 2.08(9) \\
    \end{tabular}
    \end{ruledtabular}
\end{table}

%\begin{figure}[tb]
%\begin{center}
%\includegraphics[width=0.98\columnwidth]{chi_cri_dis}
%\caption{(a) $\chi (L, \Delta_c) $  versus $L$ at several $\Delta_c$ for $n=100$. It is fitted to %the power-law $\chi (L, \Delta_c)\sim L^{\gamma/\nu}$ for all  $L\ge L_{\min}$ and the results for %the exponent $\gamma/\nu$ from different choices of $L_{\min}$ are summarized in  Table S1 of SM. %The solid lines show the power-law fits for $L_\mathrm{min} = 16$. (b) Data collapse of $\chi^*_{\rm %dis} (L, \Delta)$ with the exponent values in Table~\ref{mag_exp}.}
% \label{chi_deltac}
%\end{center}
%\end{figure}

{\it Conclusions:} We have presented a high-resolution numerical study of the 3D 3-state RFPM via a computationally efficient ground-state method that uses extrapolation in the number $n$ of initial conditions to provide quasi-exact results. With the help of FSS, all critical exponents, including the most elusive specific-heat exponent $\alpha$, are determined for finite as well infinite $n$, providing clear evidence for a {\it continuous} phase transition in the vicinity of $\Delta_c = 1.604(2)$. Given that we are working at $T=0$, it is hence clear that $3 = q < q_c^\mathrm{RF}(3)$. We are thus able to provide the first comprehensive calculation of critical exponents. As the summary in Table~\ref{tab:exponents} shows, these are close to but likely distinct from the exponents of the 3D RFIM \cite{middleton02,fytas:13}. It will be most interesting to see if this deviation grows stronger as $q$ is increased to $4$ and possibly beyond, and if and when the transition turns first order.

\begin{acknowledgements}
We thank N. Fytas for a careful reading of the manuscript. The authors are thankful to Kurt Binder for manifold discussions on topics related to the manuscript, and they dedicate this paper to his memory.
MK and MW acknowledge the support of the Royal
Society - SERB Newton International fellowship (NIF$\backslash$R1$\backslash$180386). All numerical simulations were done on the parallel compute cluster {\it Zeus} of Coventry University. The publication of this article was funded by Chemnitz University of Technology and  by the Deutsche Forschungsgemeinschaft (DFG, German Research Foundation) – 491193532.
\end{acknowledgements}

%Todo:
%\begin{itemize}
 %   \item supplementary material: plots for finite $n$, fluctuation-dissipation theorem
  %  \item energy cumulant
  %  \item legends in Fig. 3 (are different $n$ dependent or independent?)
%\end{itemize}

% The resulting plot for the best data collapse is shown in Fig.~\ref{mag_scale_rfpm}
% Fig.~\ref{mag_scale} is the scaled plot for maximum $n=100$.
% The presence of quenched random dields is known to have a profound effect on the nature of phase transition. best fit to the function of form 

\bibliography{ref.bib}

\begin{thebibliography}{65}%
\makeatletter
\providecommand \@ifxundefined [1]{%
 \@ifx{#1\undefined}
}%
\providecommand \@ifnum [1]{%
 \ifnum #1\expandafter \@firstoftwo
 \else \expandafter \@secondoftwo
 \fi
}%
\providecommand \@ifx [1]{%
 \ifx #1\expandafter \@firstoftwo
 \else \expandafter \@secondoftwo
 \fi
}%
\providecommand \natexlab [1]{#1}%
\providecommand \enquote  [1]{``#1''}%
\providecommand \bibnamefont  [1]{#1}%
\providecommand \bibfnamefont [1]{#1}%
\providecommand \citenamefont [1]{#1}%
\providecommand \href@noop [0]{\@secondoftwo}%
\providecommand \href [0]{\begingroup \@sanitize@url \@href}%
\providecommand \@href[1]{\@@startlink{#1}\@@href}%
\providecommand \@@href[1]{\endgroup#1\@@endlink}%
\providecommand \@sanitize@url [0]{\catcode `\\12\catcode `\$12\catcode
  `\&12\catcode `\#12\catcode `\^12\catcode `\_12\catcode `\%12\relax}%
\providecommand \@@startlink[1]{}%
\providecommand \@@endlink[0]{}%
\providecommand \url  [0]{\begingroup\@sanitize@url \@url }%
\providecommand \@url [1]{\endgroup\@href {#1}{\urlprefix }}%
\providecommand \urlprefix  [0]{URL }%
\providecommand \Eprint [0]{\href }%
\providecommand \doibase [0]{https://doi.org/}%
\providecommand \selectlanguage [0]{\@gobble}%
\providecommand \bibinfo  [0]{\@secondoftwo}%
\providecommand \bibfield  [0]{\@secondoftwo}%
\providecommand \translation [1]{[#1]}%
\providecommand \BibitemOpen [0]{}%
\providecommand \bibitemStop [0]{}%
\providecommand \bibitemNoStop [0]{.\EOS\space}%
\providecommand \EOS [0]{\spacefactor3000\relax}%
\providecommand \BibitemShut  [1]{\csname bibitem#1\endcsname}%
\let\auto@bib@innerbib\@empty
%</preamble>
\bibitem [{\citenamefont {Boixo}\ \emph {et~al.}(2014)\citenamefont {Boixo},
  \citenamefont {Ronnow}, \citenamefont {Isakov}, \citenamefont {Wang},
  \citenamefont {Wecker}, \citenamefont {Lidar}, \citenamefont {Martinis},\
  and\ \citenamefont {Troyer}}]{boixo:14}%
  \BibitemOpen
  \bibfield  {author} {\bibinfo {author} {\bibfnamefont {S.}~\bibnamefont
  {Boixo}}, \bibinfo {author} {\bibfnamefont {T.~F.}\ \bibnamefont {Ronnow}},
  \bibinfo {author} {\bibfnamefont {S.~V.}\ \bibnamefont {Isakov}}, \bibinfo
  {author} {\bibfnamefont {Z.}~\bibnamefont {Wang}}, \bibinfo {author}
  {\bibfnamefont {D.}~\bibnamefont {Wecker}}, \bibinfo {author} {\bibfnamefont
  {D.~A.}\ \bibnamefont {Lidar}}, \bibinfo {author} {\bibfnamefont {J.~M.}\
  \bibnamefont {Martinis}},\ and\ \bibinfo {author} {\bibfnamefont
  {M.}~\bibnamefont {Troyer}},\ }\bibfield  {title} {\bibinfo {title} {Evidence
  for quantum annealing with more than one hundred qubits},\ }\href
  {https://doi.org/10.1038/nphys2900} {\bibfield  {journal} {\bibinfo
  {journal} {Nat. Phys.}\ }\textbf {\bibinfo {volume} {10}},\ \bibinfo {pages}
  {218} (\bibinfo {year} {2014})}\BibitemShut {NoStop}%
\bibitem [{\citenamefont {Decelle}\ \emph {et~al.}(2011)\citenamefont
  {Decelle}, \citenamefont {Krzakala}, \citenamefont {Moore},\ and\
  \citenamefont {Zdeborov{\'a}}}]{decelle:11}%
  \BibitemOpen
  \bibfield  {author} {\bibinfo {author} {\bibfnamefont {A.}~\bibnamefont
  {Decelle}}, \bibinfo {author} {\bibfnamefont {F.}~\bibnamefont {Krzakala}},
  \bibinfo {author} {\bibfnamefont {C.}~\bibnamefont {Moore}},\ and\ \bibinfo
  {author} {\bibfnamefont {L.}~\bibnamefont {Zdeborov{\'a}}},\ }\bibfield
  {title} {\bibinfo {title} {Inference and phase transitions in the detection
  of modules in sparse networks},\ }\href@noop {} {\bibfield  {journal}
  {\bibinfo  {journal} {Phys. Rev. Lett.}\ }\textbf {\bibinfo {volume} {107}},\
  \bibinfo {pages} {065701} (\bibinfo {year} {2011})}\BibitemShut {NoStop}%
\bibitem [{\citenamefont {Nishimori}(2001)}]{nishimori:book}%
  \BibitemOpen
  \bibfield  {author} {\bibinfo {author} {\bibfnamefont {H.}~\bibnamefont
  {Nishimori}},\ }\href@noop {} {\emph {\bibinfo {title} {Statistical Physics
  of Spin Glasses and Information Processing}}}\ (\bibinfo  {publisher} {Oxford
  University Press},\ \bibinfo {address} {Oxford},\ \bibinfo {year}
  {2001})\BibitemShut {NoStop}%
\bibitem [{\citenamefont {Bialek}\ \emph {et~al.}(2012)\citenamefont {Bialek},
  \citenamefont {Cavagna}, \citenamefont {Giardina}, \citenamefont {Mora},
  \citenamefont {Silvestri}, \citenamefont {Viale},\ and\ \citenamefont
  {Walczak}}]{bialek:12}%
  \BibitemOpen
  \bibfield  {author} {\bibinfo {author} {\bibfnamefont {W.}~\bibnamefont
  {Bialek}}, \bibinfo {author} {\bibfnamefont {A.}~\bibnamefont {Cavagna}},
  \bibinfo {author} {\bibfnamefont {I.}~\bibnamefont {Giardina}}, \bibinfo
  {author} {\bibfnamefont {T.}~\bibnamefont {Mora}}, \bibinfo {author}
  {\bibfnamefont {E.}~\bibnamefont {Silvestri}}, \bibinfo {author}
  {\bibfnamefont {M.}~\bibnamefont {Viale}},\ and\ \bibinfo {author}
  {\bibfnamefont {A.~M.}\ \bibnamefont {Walczak}},\ }\bibfield  {title}
  {\bibinfo {title} {Statistical mechanics for natural flocks of birds},\
  }\href@noop {} {\bibfield  {journal} {\bibinfo  {journal} {PNAS}\ }\textbf
  {\bibinfo {volume} {109}},\ \bibinfo {pages} {4786} (\bibinfo {year}
  {2012})}\BibitemShut {NoStop}%
\bibitem [{\citenamefont {Young}(1997)}]{young:book}%
  \BibitemOpen
  \bibinfo {editor} {\bibfnamefont {A.~P.}\ \bibnamefont {Young}},\ ed.,\
  \href@noop {} {\emph {\bibinfo {title} {Spin Glasses and Random Fields}}}\
  (\bibinfo  {publisher} {World Scientific},\ \bibinfo {address} {Singapore},\
  \bibinfo {year} {1997})\BibitemShut {NoStop}%
\bibitem [{\citenamefont {Silevitch}\ \emph {et~al.}(2007)\citenamefont
  {Silevitch}, \citenamefont {Bitko}, \citenamefont {Brooke}, \citenamefont
  {Ghosh}, \citenamefont {Aeppli},\ and\ \citenamefont
  {Rosenbaum}}]{silevitch:07}%
  \BibitemOpen
  \bibfield  {author} {\bibinfo {author} {\bibfnamefont {D.}~\bibnamefont
  {Silevitch}}, \bibinfo {author} {\bibfnamefont {D.}~\bibnamefont {Bitko}},
  \bibinfo {author} {\bibfnamefont {J.}~\bibnamefont {Brooke}}, \bibinfo
  {author} {\bibfnamefont {S.}~\bibnamefont {Ghosh}}, \bibinfo {author}
  {\bibfnamefont {G.}~\bibnamefont {Aeppli}},\ and\ \bibinfo {author}
  {\bibfnamefont {T.}~\bibnamefont {Rosenbaum}},\ }\bibfield  {title} {\bibinfo
  {title} {A ferromagnet in a continuously tunable random field},\ }\href@noop
  {} {\bibfield  {journal} {\bibinfo  {journal} {Nature}\ }\textbf {\bibinfo
  {volume} {448}},\ \bibinfo {pages} {567} (\bibinfo {year}
  {2007})}\BibitemShut {NoStop}%
\bibitem [{\citenamefont {Xia}\ and\ \citenamefont {Wolynes}(2000)}]{xia:00}%
  \BibitemOpen
  \bibfield  {author} {\bibinfo {author} {\bibfnamefont {X.}~\bibnamefont
  {Xia}}\ and\ \bibinfo {author} {\bibfnamefont {P.~G.}\ \bibnamefont
  {Wolynes}},\ }\bibfield  {title} {\bibinfo {title} {Fragilities of liquids
  predicted from the random first order transition theory of glasses},\
  }\href@noop {} {\bibfield  {journal} {\bibinfo  {journal} {PNAS}\ }\textbf
  {\bibinfo {volume} {97}},\ \bibinfo {pages} {2990} (\bibinfo {year}
  {2000})}\BibitemShut {NoStop}%
\bibitem [{\citenamefont {Guiselin}\ \emph {et~al.}(2020)\citenamefont
  {Guiselin}, \citenamefont {Berthier},\ and\ \citenamefont
  {Tarjus}}]{guiselin:20}%
  \BibitemOpen
  \bibfield  {author} {\bibinfo {author} {\bibfnamefont {B.}~\bibnamefont
  {Guiselin}}, \bibinfo {author} {\bibfnamefont {L.}~\bibnamefont {Berthier}},\
  and\ \bibinfo {author} {\bibfnamefont {G.}~\bibnamefont {Tarjus}},\
  }\bibfield  {title} {\bibinfo {title} {Random-field {I}sing model criticality
  in a glass-forming liquid},\ }\href@noop {} {\bibfield  {journal} {\bibinfo
  {journal} {Physical Review E}\ }\textbf {\bibinfo {volume} {102}},\ \bibinfo
  {pages} {042129} (\bibinfo {year} {2020})}\BibitemShut {NoStop}%
\bibitem [{\citenamefont {Imbrie}(1984)}]{imbrie:84}%
  \BibitemOpen
  \bibfield  {author} {\bibinfo {author} {\bibfnamefont {J.~Z.}\ \bibnamefont
  {Imbrie}},\ }\bibfield  {title} {\bibinfo {title} {Lower critical dimension
  of the random-field {I}sing model},\ }\href
  {https://doi.org/doi.org/10.1103/PhysRevLett.53.1747} {\bibfield  {journal}
  {\bibinfo  {journal} {Phys. Rev. Lett.}\ }\textbf {\bibinfo {volume} {53}},\
  \bibinfo {pages} {1747} (\bibinfo {year} {1984})}\BibitemShut {NoStop}%
\bibitem [{\citenamefont {Aizenman}\ and\ \citenamefont
  {Wehr}(1989)}]{aizenmann:89a}%
  \BibitemOpen
  \bibfield  {author} {\bibinfo {author} {\bibfnamefont {M.}~\bibnamefont
  {Aizenman}}\ and\ \bibinfo {author} {\bibfnamefont {J.}~\bibnamefont
  {Wehr}},\ }\bibfield  {title} {\bibinfo {title} {Rounding of first-order
  phase transitions in systems with quenched disorder},\ }\href@noop {}
  {\bibfield  {journal} {\bibinfo  {journal} {Phys. Rev. Lett.}\ }\textbf
  {\bibinfo {volume} {62}},\ \bibinfo {pages} {2503} (\bibinfo {year}
  {1989})}\BibitemShut {NoStop}%
\bibitem [{\citenamefont {Binder}(1983)}]{binder:83}%
  \BibitemOpen
  \bibfield  {author} {\bibinfo {author} {\bibfnamefont {K.}~\bibnamefont
  {Binder}},\ }\bibfield  {title} {\bibinfo {title} {Random-field induced
  interface widths in {I}sing systems},\ }\href
  {https://doi.org/10.1007/BF01470045} {\bibfield  {journal} {\bibinfo
  {journal} {Z. Phys. B}\ }\textbf {\bibinfo {volume} {50}},\ \bibinfo {pages}
  {343} (\bibinfo {year} {1983})}\BibitemShut {NoStop}%
\bibitem [{\citenamefont {Fytas}\ and\ \citenamefont
  {Mart\'{\i}n-Mayor}(2013)}]{fytas:13}%
  \BibitemOpen
  \bibfield  {author} {\bibinfo {author} {\bibfnamefont {N.~G.}\ \bibnamefont
  {Fytas}}\ and\ \bibinfo {author} {\bibfnamefont {V.}~\bibnamefont
  {Mart\'{\i}n-Mayor}},\ }\bibfield  {title} {\bibinfo {title} {Universality in
  the three-dimensional random-field {I}sing model},\ }\href
  {https://doi.org/10.1103/PhysRevLett.110.227201} {\bibfield  {journal}
  {\bibinfo  {journal} {Phys. Rev. Lett.}\ }\textbf {\bibinfo {volume} {110}},\
  \bibinfo {pages} {227201} (\bibinfo {year} {2013})}\BibitemShut {NoStop}%
\bibitem [{\citenamefont {Fytas}\ \emph
  {et~al.}(2016{\natexlab{a}})\citenamefont {Fytas}, \citenamefont
  {Mart\'{\i}n-Mayor}, \citenamefont {Picco},\ and\ \citenamefont
  {Sourlas}}]{fytas:16}%
  \BibitemOpen
  \bibfield  {author} {\bibinfo {author} {\bibfnamefont {N.~G.}\ \bibnamefont
  {Fytas}}, \bibinfo {author} {\bibfnamefont {V.}~\bibnamefont
  {Mart\'{\i}n-Mayor}}, \bibinfo {author} {\bibfnamefont {M.}~\bibnamefont
  {Picco}},\ and\ \bibinfo {author} {\bibfnamefont {N.}~\bibnamefont
  {Sourlas}},\ }\bibfield  {title} {\bibinfo {title} {Phase transitions in
  disordered systems: the example of the random-field {I}sing model in four
  dimensions},\ }\href@noop {} {\bibfield  {journal} {\bibinfo  {journal}
  {Phys. Rev. Lett.}\ }\textbf {\bibinfo {volume} {116}},\ \bibinfo {pages}
  {227201} (\bibinfo {year} {2016}{\natexlab{a}})}\BibitemShut {NoStop}%
\bibitem [{\citenamefont {Kumar}\ \emph {et~al.}(2017)\citenamefont {Kumar},
  \citenamefont {Banerjee},\ and\ \citenamefont {Puri}}]{kumar:17}%
  \BibitemOpen
  \bibfield  {author} {\bibinfo {author} {\bibfnamefont {M.}~\bibnamefont
  {Kumar}}, \bibinfo {author} {\bibfnamefont {V.}~\bibnamefont {Banerjee}},\
  and\ \bibinfo {author} {\bibfnamefont {S.}~\bibnamefont {Puri}},\ }\bibfield
  {title} {\bibinfo {title} {Random field {I}sing model in a uniform magnetic
  field: Ground states, pinned clusters and scaling laws},\ }\href@noop {}
  {\bibfield  {journal} {\bibinfo  {journal} {Eur. Phys. J. E}\ }\textbf
  {\bibinfo {volume} {40}},\ \bibinfo {pages} {96} (\bibinfo {year}
  {2017})}\BibitemShut {NoStop}%
\bibitem [{\citenamefont {Aharony}\ \emph {et~al.}(1976)\citenamefont
  {Aharony}, \citenamefont {Imry},\ and\ \citenamefont {Ma}}]{aharony:76}%
  \BibitemOpen
  \bibfield  {author} {\bibinfo {author} {\bibfnamefont {A.}~\bibnamefont
  {Aharony}}, \bibinfo {author} {\bibfnamefont {Y.}~\bibnamefont {Imry}},\ and\
  \bibinfo {author} {\bibfnamefont {S.-k.}\ \bibnamefont {Ma}},\ }\bibfield
  {title} {\bibinfo {title} {Lowering of dimensionality in phase transitions
  with random fields},\ }\href@noop {} {\bibfield  {journal} {\bibinfo
  {journal} {Phys. Rev. Lett.}\ }\textbf {\bibinfo {volume} {37}},\ \bibinfo
  {pages} {1364} (\bibinfo {year} {1976})}\BibitemShut {NoStop}%
\bibitem [{\citenamefont {Fytas}\ \emph {et~al.}(2017)\citenamefont {Fytas},
  \citenamefont {Mart\'{\i}n-Mayor}, \citenamefont {Picco},\ and\ \citenamefont
  {Sourlas}}]{fytas:17}%
  \BibitemOpen
  \bibfield  {author} {\bibinfo {author} {\bibfnamefont {N.~G.}\ \bibnamefont
  {Fytas}}, \bibinfo {author} {\bibfnamefont {V.}~\bibnamefont
  {Mart\'{\i}n-Mayor}}, \bibinfo {author} {\bibfnamefont {M.}~\bibnamefont
  {Picco}},\ and\ \bibinfo {author} {\bibfnamefont {N.}~\bibnamefont
  {Sourlas}},\ }\bibfield  {title} {\bibinfo {title} {Restoration of
  dimensional reduction in the random-field {I}sing model at five dimensions},\
  }\href {https://doi.org/10.1103/PhysRevE.95.042117} {\bibfield  {journal}
  {\bibinfo  {journal} {Phys. Rev. E}\ }\textbf {\bibinfo {volume} {95}},\
  \bibinfo {pages} {042117} (\bibinfo {year} {2017})}\BibitemShut {NoStop}%
\bibitem [{\citenamefont {Tissier}\ and\ \citenamefont
  {Tarjus}(2011)}]{tissier:11}%
  \BibitemOpen
  \bibfield  {author} {\bibinfo {author} {\bibfnamefont {M.}~\bibnamefont
  {Tissier}}\ and\ \bibinfo {author} {\bibfnamefont {G.}~\bibnamefont
  {Tarjus}},\ }\bibfield  {title} {\bibinfo {title} {Supersymmetry and its
  spontaneous breaking in the random field {I}sing model},\ }\href
  {https://doi.org/10.1103/PhysRevLett.107.041601} {\bibfield  {journal}
  {\bibinfo  {journal} {Phys. Rev. Lett.}\ }\textbf {\bibinfo {volume} {107}},\
  \bibinfo {pages} {041601} (\bibinfo {year} {2011})}\BibitemShut {NoStop}%
\bibitem [{\citenamefont {Kaviraj}\ \emph {et~al.}(2021)\citenamefont
  {Kaviraj}, \citenamefont {Rychkov},\ and\ \citenamefont
  {Trevisani}}]{kaviraj:21}%
  \BibitemOpen
  \bibfield  {author} {\bibinfo {author} {\bibfnamefont {A.}~\bibnamefont
  {Kaviraj}}, \bibinfo {author} {\bibfnamefont {S.}~\bibnamefont {Rychkov}},\
  and\ \bibinfo {author} {\bibfnamefont {E.}~\bibnamefont {Trevisani}},\
  }\bibfield  {title} {\bibinfo {title} {The fate of {P}arisi-{S}ourlas
  supersymmetry in random field models},\ }\href@noop {} {\bibfield  {journal}
  {\bibinfo  {journal} {arXiv preprint arXiv:2112.06942}\ } (\bibinfo {year}
  {2021})}\BibitemShut {NoStop}%
\bibitem [{\citenamefont {Nishimori}(1983)}]{nishimori}%
  \BibitemOpen
  \bibfield  {author} {\bibinfo {author} {\bibfnamefont {H.}~\bibnamefont
  {Nishimori}},\ }\bibfield  {title} {\bibinfo {title} {Potts model in random
  fields},\ }\href@noop {} {\bibfield  {journal} {\bibinfo  {journal} {Phys.
  Rev. B}\ }\textbf {\bibinfo {volume} {28}},\ \bibinfo {pages} {4011}
  (\bibinfo {year} {1983})}\BibitemShut {NoStop}%
\bibitem [{\citenamefont {Blankschtein}\ \emph {et~al.}(1984)\citenamefont
  {Blankschtein}, \citenamefont {Shapir},\ and\ \citenamefont
  {Aharony}}]{shapir84}%
  \BibitemOpen
  \bibfield  {author} {\bibinfo {author} {\bibfnamefont {D.}~\bibnamefont
  {Blankschtein}}, \bibinfo {author} {\bibfnamefont {Y.}~\bibnamefont
  {Shapir}},\ and\ \bibinfo {author} {\bibfnamefont {A.}~\bibnamefont
  {Aharony}},\ }\bibfield  {title} {\bibinfo {title} {Potts models in random
  fields},\ }\href@noop {} {\bibfield  {journal} {\bibinfo  {journal} {Phys.
  Rev. B}\ }\textbf {\bibinfo {volume} {29}},\ \bibinfo {pages} {1263}
  (\bibinfo {year} {1984})}\BibitemShut {NoStop}%
\bibitem [{\citenamefont {Alford}\ \emph {et~al.}(2001)\citenamefont {Alford},
  \citenamefont {Chandrasekharan}, \citenamefont {Cox},\ and\ \citenamefont
  {Wiese}}]{alford:01}%
  \BibitemOpen
  \bibfield  {author} {\bibinfo {author} {\bibfnamefont {M.}~\bibnamefont
  {Alford}}, \bibinfo {author} {\bibfnamefont {S.}~\bibnamefont
  {Chandrasekharan}}, \bibinfo {author} {\bibfnamefont {J.}~\bibnamefont
  {Cox}},\ and\ \bibinfo {author} {\bibfnamefont {U.-J.}\ \bibnamefont
  {Wiese}},\ }\bibfield  {title} {\bibinfo {title} {Solution of the complex
  action problem in the {P}otts model for dense {QCD}},\ }\href
  {https://doi.org/https://doi.org/10.1016/S0550-3213(01)00068-2} {\bibfield
  {journal} {\bibinfo  {journal} {Nucl. Phys. B}\ }\textbf {\bibinfo {volume}
  {602}},\ \bibinfo {pages} {61} (\bibinfo {year} {2001})}\BibitemShut
  {NoStop}%
\bibitem [{\citenamefont {Domany}\ \emph {et~al.}(1982)\citenamefont {Domany},
  \citenamefont {Shnidman},\ and\ \citenamefont {Mukamel}}]{domany:82}%
  \BibitemOpen
  \bibfield  {author} {\bibinfo {author} {\bibfnamefont {E.}~\bibnamefont
  {Domany}}, \bibinfo {author} {\bibfnamefont {Y.}~\bibnamefont {Shnidman}},\
  and\ \bibinfo {author} {\bibfnamefont {D.}~\bibnamefont {Mukamel}},\
  }\bibfield  {title} {\bibinfo {title} {Type {I} {FCC} antiferromagnets in a
  magnetic field: a realisation of the \(q=3\)-and \(q=4\)-state {P}otts
  models},\ }\href@noop {} {\bibfield  {journal} {\bibinfo  {journal} {J. Phys.
  C}\ }\textbf {\bibinfo {volume} {15}},\ \bibinfo {pages} {L495} (\bibinfo
  {year} {1982})}\BibitemShut {NoStop}%
\bibitem [{\citenamefont {Binder}\ and\ \citenamefont
  {Reger}(1992)}]{binder:92}%
  \BibitemOpen
  \bibfield  {author} {\bibinfo {author} {\bibfnamefont {K.}~\bibnamefont
  {Binder}}\ and\ \bibinfo {author} {\bibfnamefont {J.}~\bibnamefont {Reger}},\
  }\bibfield  {title} {\bibinfo {title} {Theory of orientational glasses
  models, concepts, simulations},\ }\href@noop {} {\bibfield  {journal}
  {\bibinfo  {journal} {Adv. Phys.}\ }\textbf {\bibinfo {volume} {41}},\
  \bibinfo {pages} {547} (\bibinfo {year} {1992})}\BibitemShut {NoStop}%
\bibitem [{\citenamefont {Graner}\ and\ \citenamefont
  {Glazier}(1992)}]{graner:92}%
  \BibitemOpen
  \bibfield  {author} {\bibinfo {author} {\bibfnamefont {F.}~\bibnamefont
  {Graner}}\ and\ \bibinfo {author} {\bibfnamefont {J.~A.}\ \bibnamefont
  {Glazier}},\ }\bibfield  {title} {\bibinfo {title} {Simulation of biological
  cell sorting using a two-dimensional extended {P}otts model},\ }\href
  {https://doi.org/10.1103/PhysRevLett.69.2013} {\bibfield  {journal} {\bibinfo
   {journal} {Phys. Rev. Lett.}\ }\textbf {\bibinfo {volume} {69}},\ \bibinfo
  {pages} {2013} (\bibinfo {year} {1992})}\BibitemShut {NoStop}%
\bibitem [{\citenamefont {Wu}(1982)}]{wu}%
  \BibitemOpen
  \bibfield  {author} {\bibinfo {author} {\bibfnamefont {F.~Y.}\ \bibnamefont
  {Wu}},\ }\bibfield  {title} {\bibinfo {title} {The {P}otts model},\ }\href
  {https://doi.org/10.1103/RevModPhys.54.235} {\bibfield  {journal} {\bibinfo
  {journal} {Rev. Mod. Phys.}\ }\textbf {\bibinfo {volume} {54}},\ \bibinfo
  {pages} {235} (\bibinfo {year} {1982})}\BibitemShut {NoStop}%
\bibitem [{\citenamefont {Duminil-Copin}\ \emph {et~al.}(2017)\citenamefont
  {Duminil-Copin}, \citenamefont {Sidoravicius},\ and\ \citenamefont
  {Tassion}}]{duminil:15b}%
  \BibitemOpen
  \bibfield  {author} {\bibinfo {author} {\bibfnamefont {H.}~\bibnamefont
  {Duminil-Copin}}, \bibinfo {author} {\bibfnamefont {V.}~\bibnamefont
  {Sidoravicius}},\ and\ \bibinfo {author} {\bibfnamefont {V.}~\bibnamefont
  {Tassion}},\ }\bibfield  {title} {\bibinfo {title} {Continuity of the phase
  transition for planar random-cluster and {P}otts models with \(1\le q \le
  4\)},\ }\href@noop {} {\bibfield  {journal} {\bibinfo  {journal} {Commun.
  Math. Phys.}\ }\textbf {\bibinfo {volume} {349}},\ \bibinfo {pages} {47}
  (\bibinfo {year} {2017})}\BibitemShut {NoStop}%
\bibitem [{\citenamefont {Duminil-Copin}\ \emph {et~al.}(2016)\citenamefont
  {Duminil-Copin}, \citenamefont {Gagnebin}, \citenamefont {Harel},
  \citenamefont {Manolescu},\ and\ \citenamefont {Tassion}}]{duminil:16}%
  \BibitemOpen
  \bibfield  {author} {\bibinfo {author} {\bibfnamefont {H.}~\bibnamefont
  {Duminil-Copin}}, \bibinfo {author} {\bibfnamefont {M.}~\bibnamefont
  {Gagnebin}}, \bibinfo {author} {\bibfnamefont {M.}~\bibnamefont {Harel}},
  \bibinfo {author} {\bibfnamefont {I.}~\bibnamefont {Manolescu}},\ and\
  \bibinfo {author} {\bibfnamefont {V.}~\bibnamefont {Tassion}},\ }\bibfield
  {title} {\bibinfo {title} {Discontinuity of the phase transition for the
  planar random-cluster and {P}otts models with \(q>4\)},\ }\href@noop {}
  {\bibfield  {journal} {\bibinfo  {journal} {arXiv preprint arXiv:1611.09877}\
  } (\bibinfo {year} {2016})}\BibitemShut {NoStop}%
\bibitem [{\citenamefont {Hartmann}(2005)}]{hartmann:05}%
  \BibitemOpen
  \bibfield  {author} {\bibinfo {author} {\bibfnamefont {A.~K.}\ \bibnamefont
  {Hartmann}},\ }\bibfield  {title} {\bibinfo {title} {Calculation of partition
  functions by measuring component distributions},\ }\href@noop {} {\bibfield
  {journal} {\bibinfo  {journal} {Phys. Rev. Lett.}\ }\textbf {\bibinfo
  {volume} {94}},\ \bibinfo {pages} {050601} (\bibinfo {year}
  {2005})}\BibitemShut {NoStop}%
\bibitem [{\citenamefont {Cardy}(1999)}]{cardy:99a}%
  \BibitemOpen
  \bibfield  {author} {\bibinfo {author} {\bibfnamefont {J.}~\bibnamefont
  {Cardy}},\ }\bibfield  {title} {\bibinfo {title} {Quenched randomness at
  first-order transitions},\ }\href@noop {} {\bibfield  {journal} {\bibinfo
  {journal} {Physica A}\ }\textbf {\bibinfo {volume} {263}},\ \bibinfo {pages}
  {215} (\bibinfo {year} {1999})}\BibitemShut {NoStop}%
\bibitem [{\citenamefont {Kumar}\ \emph {et~al.}(2018)\citenamefont {Kumar},
  \citenamefont {Kumar}, \citenamefont {Weigel}, \citenamefont {Banerjee},
  \citenamefont {Janke},\ and\ \citenamefont {Puri}}]{kumar2018approximate}%
  \BibitemOpen
  \bibfield  {author} {\bibinfo {author} {\bibfnamefont {M.}~\bibnamefont
  {Kumar}}, \bibinfo {author} {\bibfnamefont {R.}~\bibnamefont {Kumar}},
  \bibinfo {author} {\bibfnamefont {M.}~\bibnamefont {Weigel}}, \bibinfo
  {author} {\bibfnamefont {V.}~\bibnamefont {Banerjee}}, \bibinfo {author}
  {\bibfnamefont {W.}~\bibnamefont {Janke}},\ and\ \bibinfo {author}
  {\bibfnamefont {S.}~\bibnamefont {Puri}},\ }\bibfield  {title} {\bibinfo
  {title} {Approximate ground states of the random-field {P}otts model from
  graph cuts},\ }\href@noop {} {\bibfield  {journal} {\bibinfo  {journal}
  {Phys. Rev. E}\ }\textbf {\bibinfo {volume} {97}},\ \bibinfo {pages} {053307}
  (\bibinfo {year} {2018})}\BibitemShut {NoStop}%
\bibitem [{\citenamefont {Eichhorn}\ and\ \citenamefont
  {Binder}(1996)}]{ebjpcm}%
  \BibitemOpen
  \bibfield  {author} {\bibinfo {author} {\bibfnamefont {K.}~\bibnamefont
  {Eichhorn}}\ and\ \bibinfo {author} {\bibfnamefont {K.}~\bibnamefont
  {Binder}},\ }\bibfield  {title} {\bibinfo {title} {Monte {C}arlo
  investigation of the three-dimensional random-field three-state {P}otts
  model},\ }\href@noop {} {\bibfield  {journal} {\bibinfo  {journal} {J. Phys.:
  Condens. Matter}\ }\textbf {\bibinfo {volume} {8}},\ \bibinfo {pages} {5209}
  (\bibinfo {year} {1996})}\BibitemShut {NoStop}%
\bibitem [{\citenamefont {Eichhorn}\ and\ \citenamefont
  {Binder}(1995{\natexlab{a}})}]{eb_epl}%
  \BibitemOpen
  \bibfield  {author} {\bibinfo {author} {\bibfnamefont {K.}~\bibnamefont
  {Eichhorn}}\ and\ \bibinfo {author} {\bibfnamefont {K.}~\bibnamefont
  {Binder}},\ }\bibfield  {title} {\bibinfo {title} {Finite-size scaling study
  of the three-state {P}otts model in random fields: Evidence for a
  second-order transition},\ }\href@noop {} {\bibfield  {journal} {\bibinfo
  {journal} {Europhys. Lett.}\ }\textbf {\bibinfo {volume} {30}},\ \bibinfo
  {pages} {331} (\bibinfo {year} {1995}{\natexlab{a}})}\BibitemShut {NoStop}%
\bibitem [{\citenamefont {Goldschmidt}\ and\ \citenamefont
  {Xu}(1986)}]{goldschmidt:86}%
  \BibitemOpen
  \bibfield  {author} {\bibinfo {author} {\bibfnamefont {Y.~Y.}\ \bibnamefont
  {Goldschmidt}}\ and\ \bibinfo {author} {\bibfnamefont {G.}~\bibnamefont
  {Xu}},\ }\bibfield  {title} {\bibinfo {title} {The random-field potts model
  in three dimensions},\ }\href@noop {} {\bibfield  {journal} {\bibinfo
  {journal} {Nucl. Phys. B}\ }\textbf {\bibinfo {volume} {265}},\ \bibinfo
  {pages} {1} (\bibinfo {year} {1986})}\BibitemShut {NoStop}%
\bibitem [{\citenamefont {Eichhorn}\ and\ \citenamefont
  {Binder}(1995{\natexlab{b}})}]{ebzpb}%
  \BibitemOpen
  \bibfield  {author} {\bibinfo {author} {\bibfnamefont {K.}~\bibnamefont
  {Eichhorn}}\ and\ \bibinfo {author} {\bibfnamefont {K.}~\bibnamefont
  {Binder}},\ }\bibfield  {title} {\bibinfo {title} {The three-dimensional
  three-state {P}otts ferromagnet exposed to random fields: evidence for a
  second order transition},\ }\href@noop {} {\bibfield  {journal} {\bibinfo
  {journal} {Z. Phys. B}\ }\textbf {\bibinfo {volume} {99}},\ \bibinfo {pages}
  {413} (\bibinfo {year} {1995}{\natexlab{b}})}\BibitemShut {NoStop}%
\bibitem [{\citenamefont {Türkoğlu}\ and\ \citenamefont
  {Berker}(2021)}]{turkoglu:21}%
  \BibitemOpen
  \bibfield  {author} {\bibinfo {author} {\bibfnamefont {A.}~\bibnamefont
  {Türkoğlu}}\ and\ \bibinfo {author} {\bibfnamefont {A.~N.}\ \bibnamefont
  {Berker}},\ }\bibfield  {title} {\bibinfo {title} {Phase transitions of the
  variety of random-field {P}otts models},\ }\href
  {https://doi.org/https://doi.org/10.1016/j.physa.2021.126339} {\bibfield
  {journal} {\bibinfo  {journal} {Physica A}\ }\textbf {\bibinfo {volume}
  {583}},\ \bibinfo {pages} {126339} (\bibinfo {year} {2021})}\BibitemShut
  {NoStop}%
\bibitem [{\citenamefont {Aharony}\ \emph {et~al.}(1977)\citenamefont
  {Aharony}, \citenamefont {M{\"u}ller},\ and\ \citenamefont
  {Berlinger}}]{aharony}%
  \BibitemOpen
  \bibfield  {author} {\bibinfo {author} {\bibfnamefont {A.}~\bibnamefont
  {Aharony}}, \bibinfo {author} {\bibfnamefont {K.}~\bibnamefont
  {M{\"u}ller}},\ and\ \bibinfo {author} {\bibfnamefont {W.}~\bibnamefont
  {Berlinger}},\ }\bibfield  {title} {\bibinfo {title} {Trigonal-to-tetragonal
  transition in stressed {SrTiO}$_3$: A realization of the three-state {P}otts
  model},\ }\href@noop {} {\bibfield  {journal} {\bibinfo  {journal} {Phys.
  Rev. Lett.}\ }\textbf {\bibinfo {volume} {38}},\ \bibinfo {pages} {33}
  (\bibinfo {year} {1977})}\BibitemShut {NoStop}%
\bibitem [{\citenamefont {Mukamel}(1981)}]{mukamel1981phase}%
  \BibitemOpen
  \bibfield  {author} {\bibinfo {author} {\bibfnamefont {D.}~\bibnamefont
  {Mukamel}},\ }\bibfield  {title} {\bibinfo {title} {Phase diagrams and
  multicritical points in randomly mixed alloys},\ }\href@noop {} {\bibfield
  {journal} {\bibinfo  {journal} {Phys. Rev. Lett.}\ }\textbf {\bibinfo
  {volume} {46}},\ \bibinfo {pages} {845} (\bibinfo {year} {1981})}\BibitemShut
  {NoStop}%
\bibitem [{\citenamefont {Reed}(1985)}]{reed}%
  \BibitemOpen
  \bibfield  {author} {\bibinfo {author} {\bibfnamefont {P.}~\bibnamefont
  {Reed}},\ }\bibfield  {title} {\bibinfo {title} {The {P}otts model in a
  random field: a {M}onte {C}arlo study},\ }\href@noop {} {\bibfield  {journal}
  {\bibinfo  {journal} {J. Phys. C: Solid State Phys.}\ }\textbf {\bibinfo
  {volume} {18}},\ \bibinfo {pages} {L615} (\bibinfo {year}
  {1985})}\BibitemShut {NoStop}%
\bibitem [{\citenamefont {Goldschmidt}\ and\ \citenamefont {Xu}(1985)}]{gx}%
  \BibitemOpen
  \bibfield  {author} {\bibinfo {author} {\bibfnamefont {Y.~Y.}\ \bibnamefont
  {Goldschmidt}}\ and\ \bibinfo {author} {\bibfnamefont {G.}~\bibnamefont
  {Xu}},\ }\bibfield  {title} {\bibinfo {title} {Phase diagrams of the
  random-field {P}otts model in three dimensions},\ }\href@noop {} {\bibfield
  {journal} {\bibinfo  {journal} {Phys. Rev. B}\ }\textbf {\bibinfo {volume}
  {32}},\ \bibinfo {pages} {1876} (\bibinfo {year} {1985})}\BibitemShut
  {NoStop}%
\bibitem [{\citenamefont {d'Auriac}\ \emph {et~al.}(1985)\citenamefont
  {d'Auriac}, \citenamefont {Preissmann},\ and\ \citenamefont
  {Rammal}}]{dauriac:85}%
  \BibitemOpen
  \bibfield  {author} {\bibinfo {author} {\bibfnamefont {J.~A.}\ \bibnamefont
  {d'Auriac}}, \bibinfo {author} {\bibfnamefont {M.}~\bibnamefont
  {Preissmann}},\ and\ \bibinfo {author} {\bibfnamefont {R.}~\bibnamefont
  {Rammal}},\ }\bibfield  {title} {\bibinfo {title} {The random field {I}sing
  model: algorithmic complexity and phase transition},\ }\href@noop {}
  {\bibfield  {journal} {\bibinfo  {journal} {J. Phys. Lett.}\ }\textbf
  {\bibinfo {volume} {46}},\ \bibinfo {pages} {173} (\bibinfo {year}
  {1985})}\BibitemShut {NoStop}%
\bibitem [{\citenamefont {Middleton}\ and\ \citenamefont
  {Fisher}(2002)}]{middleton02}%
  \BibitemOpen
  \bibfield  {author} {\bibinfo {author} {\bibfnamefont {A.~A.}\ \bibnamefont
  {Middleton}}\ and\ \bibinfo {author} {\bibfnamefont {D.~S.}\ \bibnamefont
  {Fisher}},\ }\bibfield  {title} {\bibinfo {title} {Three-dimensional
  random-field {I}sing magnet: Interfaces, scaling, and the nature of states},\
  }\href@noop {} {\bibfield  {journal} {\bibinfo  {journal} {Phys. Rev. B}\
  }\textbf {\bibinfo {volume} {65}},\ \bibinfo {pages} {134411} (\bibinfo
  {year} {2002})}\BibitemShut {NoStop}%
\bibitem [{\citenamefont {Stevenson}\ and\ \citenamefont
  {Weigel}(2011)}]{stevenson:11}%
  \BibitemOpen
  \bibfield  {author} {\bibinfo {author} {\bibfnamefont {J.~D.}\ \bibnamefont
  {Stevenson}}\ and\ \bibinfo {author} {\bibfnamefont {M.}~\bibnamefont
  {Weigel}},\ }\bibfield  {title} {\bibinfo {title} {Domain walls and
  {Schramm-Loewner} evolution in the random-field {I}sing model},\ }\href@noop
  {} {\bibfield  {journal} {\bibinfo  {journal} {Europhys. Lett.}\ }\textbf
  {\bibinfo {volume} {95}},\ \bibinfo {pages} {40001} (\bibinfo {year}
  {2011})}\BibitemShut {NoStop}%
\bibitem [{\citenamefont {Boykov}\ \emph {et~al.}(2001)\citenamefont {Boykov},
  \citenamefont {Veksler},\ and\ \citenamefont {Zabih}}]{bvz}%
  \BibitemOpen
  \bibfield  {author} {\bibinfo {author} {\bibfnamefont {Y.}~\bibnamefont
  {Boykov}}, \bibinfo {author} {\bibfnamefont {O.}~\bibnamefont {Veksler}},\
  and\ \bibinfo {author} {\bibfnamefont {R.}~\bibnamefont {Zabih}},\ }\bibfield
   {title} {\bibinfo {title} {Fast approximate energy minimization via graph
  cuts},\ }\href@noop {} {\bibfield  {journal} {\bibinfo  {journal} {IEEE
  Trans. Pattern Anal. Mach. Intell.}\ }\textbf {\bibinfo {volume} {23}},\
  \bibinfo {pages} {1222} (\bibinfo {year} {2001})}\BibitemShut {NoStop}%
\bibitem [{\citenamefont {Kolmogorov}\ and\ \citenamefont
  {Zabin}(2004)}]{kolmogorov2004energy}%
  \BibitemOpen
  \bibfield  {author} {\bibinfo {author} {\bibfnamefont {V.}~\bibnamefont
  {Kolmogorov}}\ and\ \bibinfo {author} {\bibfnamefont {R.}~\bibnamefont
  {Zabin}},\ }\bibfield  {title} {\bibinfo {title} {What energy functions can
  be minimized via graph cuts?},\ }\href@noop {} {\bibfield  {journal}
  {\bibinfo  {journal} {IEEE Trans. Pattern Anal. Mach. Intell.}\ }\textbf
  {\bibinfo {volume} {26}},\ \bibinfo {pages} {147} (\bibinfo {year}
  {2004})}\BibitemShut {NoStop}%
\bibitem [{\citenamefont {Boykov}\ and\ \citenamefont
  {Kolmogorov}(2004)}]{kolmogorov:04}%
  \BibitemOpen
  \bibfield  {author} {\bibinfo {author} {\bibfnamefont {Y.}~\bibnamefont
  {Boykov}}\ and\ \bibinfo {author} {\bibfnamefont {V.}~\bibnamefont
  {Kolmogorov}},\ }\bibfield  {title} {\bibinfo {title} {An experimental
  comparison of min-cut/max-flow algorithms for energy minimization in
  vision},\ }\href@noop {} {\bibfield  {journal} {\bibinfo  {journal} {IEEE
  Trans. Pattern Anal. Mach. Intell.}\ }\textbf {\bibinfo {volume} {26}},\
  \bibinfo {pages} {1124} (\bibinfo {year} {2004})}\BibitemShut {NoStop}%
\bibitem [{\citenamefont {Kumar}\ and\ \citenamefont
  {Weigel}(2022)}]{extrapolate}%
  \BibitemOpen
  \bibfield  {author} {\bibinfo {author} {\bibfnamefont {M.}~\bibnamefont
  {Kumar}}\ and\ \bibinfo {author} {\bibfnamefont {M.}~\bibnamefont {Weigel}},\
  }\bibfield  {title} {\bibinfo {title} {Quasi-exact ground-state algorithm for
  the random-field {P}otts model},\ }\href@noop {} {\bibfield  {journal}
  {\bibinfo  {journal} {arXiv preprint arXiv:2204.11745}\ } (\bibinfo {year}
  {2022})}\BibitemShut {NoStop}%
\bibitem [{\citenamefont {Hartmann}\ and\ \citenamefont
  {Young}(2001)}]{hartmann:01b}%
  \BibitemOpen
  \bibfield  {author} {\bibinfo {author} {\bibfnamefont {A.}~\bibnamefont
  {Hartmann}}\ and\ \bibinfo {author} {\bibfnamefont {A.}~\bibnamefont
  {Young}},\ }\bibfield  {title} {\bibinfo {title} {Specific-heat exponent of
  random-field systems via ground-state calculations},\ }\href@noop {}
  {\bibfield  {journal} {\bibinfo  {journal} {Phys. Rev. B}\ }\textbf {\bibinfo
  {volume} {64}},\ \bibinfo {pages} {214419} (\bibinfo {year}
  {2001})}\BibitemShut {NoStop}%
\bibitem [{\citenamefont {Efron}(1982)}]{efron82}%
  \BibitemOpen
  \bibfield  {author} {\bibinfo {author} {\bibfnamefont {B.}~\bibnamefont
  {Efron}},\ }\href@noop {} {\emph {\bibinfo {title} {The jackknife, the
  bootstrap and other resampling plans}}}\ (\bibinfo  {publisher} {SIAM},\
  \bibinfo {year} {1982})\BibitemShut {NoStop}%
\bibitem [{\citenamefont {Miller}(1974)}]{miller74}%
  \BibitemOpen
  \bibfield  {author} {\bibinfo {author} {\bibfnamefont {R.~G.}\ \bibnamefont
  {Miller}},\ }\bibfield  {title} {\bibinfo {title} {The jackknife-a review},\
  }\href@noop {} {\bibfield  {journal} {\bibinfo  {journal} {Biometrika}\
  }\textbf {\bibinfo {volume} {61}},\ \bibinfo {pages} {1} (\bibinfo {year}
  {1974})}\BibitemShut {NoStop}%
\bibitem [{\citenamefont {Weigel}\ and\ \citenamefont
  {Janke}(2010)}]{weigel:10}%
  \BibitemOpen
  \bibfield  {author} {\bibinfo {author} {\bibfnamefont {M.}~\bibnamefont
  {Weigel}}\ and\ \bibinfo {author} {\bibfnamefont {W.}~\bibnamefont {Janke}},\
  }\bibfield  {title} {\bibinfo {title} {Error estimation and reduction with
  cross correlations},\ }\href {https://doi.org/10.1103/PhysRevE.81.066701}
  {\bibfield  {journal} {\bibinfo  {journal} {Phys. Rev. E}\ }\textbf {\bibinfo
  {volume} {81}},\ \bibinfo {pages} {066701} (\bibinfo {year}
  {2010})}\BibitemShut {NoStop}%
\bibitem [{\citenamefont {Melchert}(2009)}]{melchert09}%
  \BibitemOpen
  \bibfield  {author} {\bibinfo {author} {\bibfnamefont {O.}~\bibnamefont
  {Melchert}},\ }\bibfield  {title} {\bibinfo {title} {autoscale.py --- a
  program for automatic finite-size scaling analyses: A user's guide},\
  }\href@noop {} {\bibfield  {journal} {\bibinfo  {journal} {arXiv preprint
  arXiv:0910.5403}\ } (\bibinfo {year} {2009})}\BibitemShut {NoStop}%
\bibitem [{Note1()}]{Note1}%
  \BibitemOpen
  \bibinfo {note} {We have explicitly verified that a slight variation in the
  exponent $b$ does not yield significantly different results in the
  extrapolations.}\BibitemShut {Stop}%
\bibitem [{SM()}]{SM}%
  \BibitemOpen
  \href@noop {} {\ }\bibinfo {note} {See Supplemental Material at [URL will be
  inserted by publisher] for details regarding the extrapolation for physical
  quantities, the order of the transition, and the fluctuation formula for the
  susceptibility.}\BibitemShut {Stop}%
\bibitem [{\citenamefont {Privman}(1990)}]{privman:privman}%
  \BibitemOpen
  \bibfield  {author} {\bibinfo {author} {\bibfnamefont {V.}~\bibnamefont
  {Privman}},\ }\bibfield  {title} {\bibinfo {title} {Finite-size scaling
  theory},\ }in\ \href@noop {} {\emph {\bibinfo {booktitle} {Finite Size
  Scaling and Numerical Simulation of Statistical Systems}}},\ \bibinfo
  {editor} {edited by\ \bibinfo {editor} {\bibfnamefont {V.}~\bibnamefont
  {Privman}}}\ (\bibinfo  {publisher} {World Scientific},\ \bibinfo {address}
  {Singapore},\ \bibinfo {year} {1990})\ pp.\ \bibinfo {pages}
  {1--98}\BibitemShut {NoStop}%
\bibitem [{\citenamefont {Houdayer}\ and\ \citenamefont
  {Hartmann}(2004)}]{houdayer:04}%
  \BibitemOpen
  \bibfield  {author} {\bibinfo {author} {\bibfnamefont {J.}~\bibnamefont
  {Houdayer}}\ and\ \bibinfo {author} {\bibfnamefont {A.~K.}\ \bibnamefont
  {Hartmann}},\ }\bibfield  {title} {\bibinfo {title} {Low-temperature behavior
  of two-dimensional gaussian {I}sing spin glasses},\ }\href
  {https://doi.org/10.1103/PhysRevB.70.014418} {\bibfield  {journal} {\bibinfo
  {journal} {Phys. Rev. B}\ }\textbf {\bibinfo {volume} {70}},\ \bibinfo
  {pages} {014418} (\bibinfo {year} {2004})}\BibitemShut {NoStop}%
\bibitem [{\citenamefont {Binder}(1981)}]{binder81}%
  \BibitemOpen
  \bibfield  {author} {\bibinfo {author} {\bibfnamefont {K.}~\bibnamefont
  {Binder}},\ }\bibfield  {title} {\bibinfo {title} {Finite size scaling
  analysis of {I}sing model block distribution functions},\ }\href@noop {}
  {\bibfield  {journal} {\bibinfo  {journal} {Z. Phys.}\ }\textbf {\bibinfo
  {volume} {43}},\ \bibinfo {pages} {119} (\bibinfo {year} {1981})}\BibitemShut
  {NoStop}%
\bibitem [{\citenamefont {Binder}\ and\ \citenamefont
  {Heermann}(2010)}]{binder:book1}%
  \BibitemOpen
  \bibfield  {author} {\bibinfo {author} {\bibfnamefont {K.}~\bibnamefont
  {Binder}}\ and\ \bibinfo {author} {\bibfnamefont {D.~W.}\ \bibnamefont
  {Heermann}},\ }\href@noop {} {\emph {\bibinfo {title} {Monte Carlo Simulation
  in Statistical Physics}}},\ \bibinfo {edition} {5th}\ ed.\ (\bibinfo
  {publisher} {Springer-Verlag},\ \bibinfo {address} {Berlin Heidelberg},\
  \bibinfo {year} {2010})\BibitemShut {NoStop}%
\bibitem [{\citenamefont {Fytas}\ \emph
  {et~al.}(2016{\natexlab{b}})\citenamefont {Fytas}, \citenamefont
  {Theodorakis},\ and\ \citenamefont {Hartmann}}]{fytas:16b}%
  \BibitemOpen
  \bibfield  {author} {\bibinfo {author} {\bibfnamefont {N.~G.}\ \bibnamefont
  {Fytas}}, \bibinfo {author} {\bibfnamefont {P.~E.}\ \bibnamefont
  {Theodorakis}},\ and\ \bibinfo {author} {\bibfnamefont {A.~K.}\ \bibnamefont
  {Hartmann}},\ }\bibfield  {title} {\bibinfo {title} {Revisiting the scaling
  of the specific heat of the three-dimensional random-field {I}sing model},\
  }\href@noop {} {\bibfield  {journal} {\bibinfo  {journal} {Eur. Phys. J. B}\
  }\textbf {\bibinfo {volume} {89}},\ \bibinfo {pages} {1} (\bibinfo {year}
  {2016}{\natexlab{b}})}\BibitemShut {NoStop}%
\bibitem [{\citenamefont {Press}\ \emph {et~al.}(2007)\citenamefont {Press},
  \citenamefont {Teukolsky}, \citenamefont {Vetterling},\ and\ \citenamefont
  {Flannery}}]{numrec}%
  \BibitemOpen
  \bibfield  {author} {\bibinfo {author} {\bibfnamefont {W.~H.}\ \bibnamefont
  {Press}}, \bibinfo {author} {\bibfnamefont {S.~A.}\ \bibnamefont
  {Teukolsky}}, \bibinfo {author} {\bibfnamefont {W.~T.}\ \bibnamefont
  {Vetterling}},\ and\ \bibinfo {author} {\bibfnamefont {B.~P.}\ \bibnamefont
  {Flannery}},\ }\href@noop {} {\emph {\bibinfo {title} {Numerical Recipes: The
  Art of Scientific Computing}}},\ \bibinfo {edition} {3rd}\ ed.\ (\bibinfo
  {publisher} {Cambridge University Press},\ \bibinfo {address} {Cambridge},\
  \bibinfo {year} {2007})\BibitemShut {NoStop}%
\bibitem [{\citenamefont {Schwartz}\ and\ \citenamefont
  {Soffer}(1985)}]{schwartz1985exact}%
  \BibitemOpen
  \bibfield  {author} {\bibinfo {author} {\bibfnamefont {M.}~\bibnamefont
  {Schwartz}}\ and\ \bibinfo {author} {\bibfnamefont {A.}~\bibnamefont
  {Soffer}},\ }\bibfield  {title} {\bibinfo {title} {Exact inequality for
  random systems: application to random fields},\ }\href@noop {} {\bibfield
  {journal} {\bibinfo  {journal} {Phys. Rev. Lett.}\ }\textbf {\bibinfo
  {volume} {55}},\ \bibinfo {pages} {2499} (\bibinfo {year}
  {1985})}\BibitemShut {NoStop}%
\bibitem [{\citenamefont {Fisher}(1986)}]{fisher1986scaling}%
  \BibitemOpen
  \bibfield  {author} {\bibinfo {author} {\bibfnamefont {D.~S.}\ \bibnamefont
  {Fisher}},\ }\bibfield  {title} {\bibinfo {title} {Scaling and critical
  slowing down in random-field {I}sing systems},\ }\href@noop {} {\bibfield
  {journal} {\bibinfo  {journal} {Phys. Rev. Lett.}\ }\textbf {\bibinfo
  {volume} {56}},\ \bibinfo {pages} {416} (\bibinfo {year} {1986})}\BibitemShut
  {NoStop}%
\bibitem [{\citenamefont {Schwartz}(1985)}]{schwartz1985random}%
  \BibitemOpen
  \bibfield  {author} {\bibinfo {author} {\bibfnamefont {M.}~\bibnamefont
  {Schwartz}},\ }\bibfield  {title} {\bibinfo {title} {The random-field puzzle.
  {I.} {S}olution by equivalent annealing},\ }\href@noop {} {\bibfield
  {journal} {\bibinfo  {journal} {J. Phys. C: Solid State Phys.}\ }\textbf
  {\bibinfo {volume} {18}},\ \bibinfo {pages} {135} (\bibinfo {year}
  {1985})}\BibitemShut {NoStop}%
\bibitem [{\citenamefont {Schwartz}\ \emph {et~al.}(1991)\citenamefont
  {Schwartz}, \citenamefont {Gofman},\ and\ \citenamefont
  {Natterman}}]{schwartz1991missing}%
  \BibitemOpen
  \bibfield  {author} {\bibinfo {author} {\bibfnamefont {M.}~\bibnamefont
  {Schwartz}}, \bibinfo {author} {\bibfnamefont {M.}~\bibnamefont {Gofman}},\
  and\ \bibinfo {author} {\bibfnamefont {T.}~\bibnamefont {Natterman}},\
  }\bibfield  {title} {\bibinfo {title} {On the missing scaling relation in
  random field systems},\ }\href@noop {} {\bibfield  {journal} {\bibinfo
  {journal} {Physica A}\ }\textbf {\bibinfo {volume} {178}},\ \bibinfo {pages}
  {6} (\bibinfo {year} {1991})}\BibitemShut {NoStop}%
\bibitem [{\citenamefont {Schwartz}(1991)}]{schwartz1991breakdown}%
  \BibitemOpen
  \bibfield  {author} {\bibinfo {author} {\bibfnamefont {M.}~\bibnamefont
  {Schwartz}},\ }\bibfield  {title} {\bibinfo {title} {Breakdown of
  hyperscaling in random systems --- an inequality},\ }\href@noop {} {\bibfield
   {journal} {\bibinfo  {journal} {Europhys. Lett.}\ }\textbf {\bibinfo
  {volume} {15}},\ \bibinfo {pages} {777} (\bibinfo {year} {1991})}\BibitemShut
  {NoStop}%
\bibitem [{\citenamefont {Grinstein}(1976)}]{grinstein1976ferromagnetic}%
  \BibitemOpen
  \bibfield  {author} {\bibinfo {author} {\bibfnamefont {G.}~\bibnamefont
  {Grinstein}},\ }\bibfield  {title} {\bibinfo {title} {Ferromagnetic phase
  transitions in random fields: the breakdown of scaling laws},\ }\href@noop {}
  {\bibfield  {journal} {\bibinfo  {journal} {Phys. Rev. Lett.}\ }\textbf
  {\bibinfo {volume} {37}},\ \bibinfo {pages} {944} (\bibinfo {year}
  {1976})}\BibitemShut {NoStop}%
\end{thebibliography}%


\begin{thebibliography}{4}%
\makeatletter
\providecommand \@ifxundefined [1]{%
 \@ifx{#1\undefined}
}%
\providecommand \@ifnum [1]{%
 \ifnum #1\expandafter \@firstoftwo
 \else \expandafter \@secondoftwo
 \fi
}%
\providecommand \@ifx [1]{%
 \ifx #1\expandafter \@firstoftwo
 \else \expandafter \@secondoftwo
 \fi
}%
\providecommand \natexlab [1]{#1}%
\providecommand \enquote  [1]{``#1''}%
\providecommand \bibnamefont  [1]{#1}%
\providecommand \bibfnamefont [1]{#1}%
\providecommand \citenamefont [1]{#1}%
\providecommand \href@noop [0]{\@secondoftwo}%
\providecommand \href [0]{\begingroup \@sanitize@url \@href}%
\providecommand \@href[1]{\@@startlink{#1}\@@href}%
\providecommand \@@href[1]{\endgroup#1\@@endlink}%
\providecommand \@sanitize@url [0]{\catcode `\\12\catcode `\$12\catcode
  `\&12\catcode `\#12\catcode `\^12\catcode `\_12\catcode `\%12\relax}%
\providecommand \@@startlink[1]{}%
\providecommand \@@endlink[0]{}%
\providecommand \url  [0]{\begingroup\@sanitize@url \@url }%
\providecommand \@url [1]{\endgroup\@href {#1}{\urlprefix }}%
\providecommand \urlprefix  [0]{URL }%
\providecommand \Eprint [0]{\href }%
\providecommand \doibase [0]{https://doi.org/}%
\providecommand \selectlanguage [0]{\@gobble}%
\providecommand \bibinfo  [0]{\@secondoftwo}%
\providecommand \bibfield  [0]{\@secondoftwo}%
\providecommand \translation [1]{[#1]}%
\providecommand \BibitemOpen [0]{}%
\providecommand \bibitemStop [0]{}%
\providecommand \bibitemNoStop [0]{.\EOS\space}%
\providecommand \EOS [0]{\spacefactor3000\relax}%
\providecommand \BibitemShut  [1]{\csname bibitem#1\endcsname}%
\let\auto@bib@innerbib\@empty
%</preamble>
\bibitem [{\citenamefont {Challa}\ \emph {et~al.}(1986)\citenamefont {Challa},
  \citenamefont {Landau},\ and\ \citenamefont {Binder}}]{challa:86}%
  \BibitemOpen
  \bibfield  {author} {\bibinfo {author} {\bibfnamefont {M.~S.~S.}\
  \bibnamefont {Challa}}, \bibinfo {author} {\bibfnamefont {D.~P.}\
  \bibnamefont {Landau}},\ and\ \bibinfo {author} {\bibfnamefont
  {K.}~\bibnamefont {Binder}},\ }\bibfield  {title} {\bibinfo {title}
  {Finite-size effects at temperature-driven first-order transitions},\ }\href
  {https://doi.org/10.1103/physrevb.34.1841} {\bibfield  {journal} {\bibinfo
  {journal} {Phys. Rev. B}\ }\textbf {\bibinfo {volume} {34}},\ \bibinfo
  {pages} {1841} (\bibinfo {year} {1986})}\BibitemShut {NoStop}%
\bibitem [{\citenamefont {Binder}\ and\ \citenamefont
  {Heermann}(2010)}]{binder:book1}%
  \BibitemOpen
  \bibfield  {author} {\bibinfo {author} {\bibfnamefont {K.}~\bibnamefont
  {Binder}}\ and\ \bibinfo {author} {\bibfnamefont {D.~W.}\ \bibnamefont
  {Heermann}},\ }\href@noop {} {\emph {\bibinfo {title} {Monte Carlo Simulation
  in Statistical Physics}}},\ \bibinfo {edition} {5th}\ ed.\ (\bibinfo
  {publisher} {Springer-Verlag},\ \bibinfo {address} {Berlin Heidelberg},\
  \bibinfo {year} {2010})\BibitemShut {NoStop}%
\bibitem [{\citenamefont {Schwartz}\ and\ \citenamefont
  {Soffer}(1985)}]{schwartz1985exact}%
  \BibitemOpen
  \bibfield  {author} {\bibinfo {author} {\bibfnamefont {M.}~\bibnamefont
  {Schwartz}}\ and\ \bibinfo {author} {\bibfnamefont {A.}~\bibnamefont
  {Soffer}},\ }\bibfield  {title} {\bibinfo {title} {Exact inequality for
  random systems: application to random fields},\ }\href@noop {} {\bibfield
  {journal} {\bibinfo  {journal} {Phys. Rev. Lett.}\ }\textbf {\bibinfo
  {volume} {55}},\ \bibinfo {pages} {2499} (\bibinfo {year}
  {1985})}\BibitemShut {NoStop}%
\bibitem [{\citenamefont {Ahrens}\ and\ \citenamefont
  {Hartmann}(2011)}]{ahrens:11a}%
  \BibitemOpen
  \bibfield  {author} {\bibinfo {author} {\bibfnamefont {B.}~\bibnamefont
  {Ahrens}}\ and\ \bibinfo {author} {\bibfnamefont {A.~K.}\ \bibnamefont
  {Hartmann}},\ }\bibfield  {title} {\bibinfo {title} {Critical behavior of the
  random-field {I}sing model at and beyond the upper critical dimension},\
  }\href@noop {} {\bibfield  {journal} {\bibinfo  {journal} {Phys. Rev. B}\
  }\textbf {\bibinfo {volume} {83}},\ \bibinfo {pages} {014205} (\bibinfo
  {year} {2011})}\BibitemShut {NoStop}%
\end{thebibliography}%

\end{document}

\title{Supplementary Material: Critical behavior of the three-state random-field Potts model in three dimensions}
 
 \author{Manoj Kumar$^{1}$, Varsha Banerjee$^{2}$, Sanjay Puri$^{3}$, and Martin Weigel$^{1}$}
\affiliation{$^{1}$Institut für Physik, Technische Universität Chemnitz, 09107 Chemnitz, Germany.\\$^{2}$Department of Physics, Indian Institute of Technology, Hauz Khas, New Delhi -- 110016, India.\\$^{3}$School of Physical Sciences, Jawaharlal Nehru University, New Delhi -- 110067, India.}

\date{\today}

%\pacs{75.50.Lk, 64.60.F-, 02.60.Pn}
\maketitle

\section{Extrapolation of physical quantities}
\begin{figure}
\begin{center}
\includegraphics[width=0.95\linewidth]{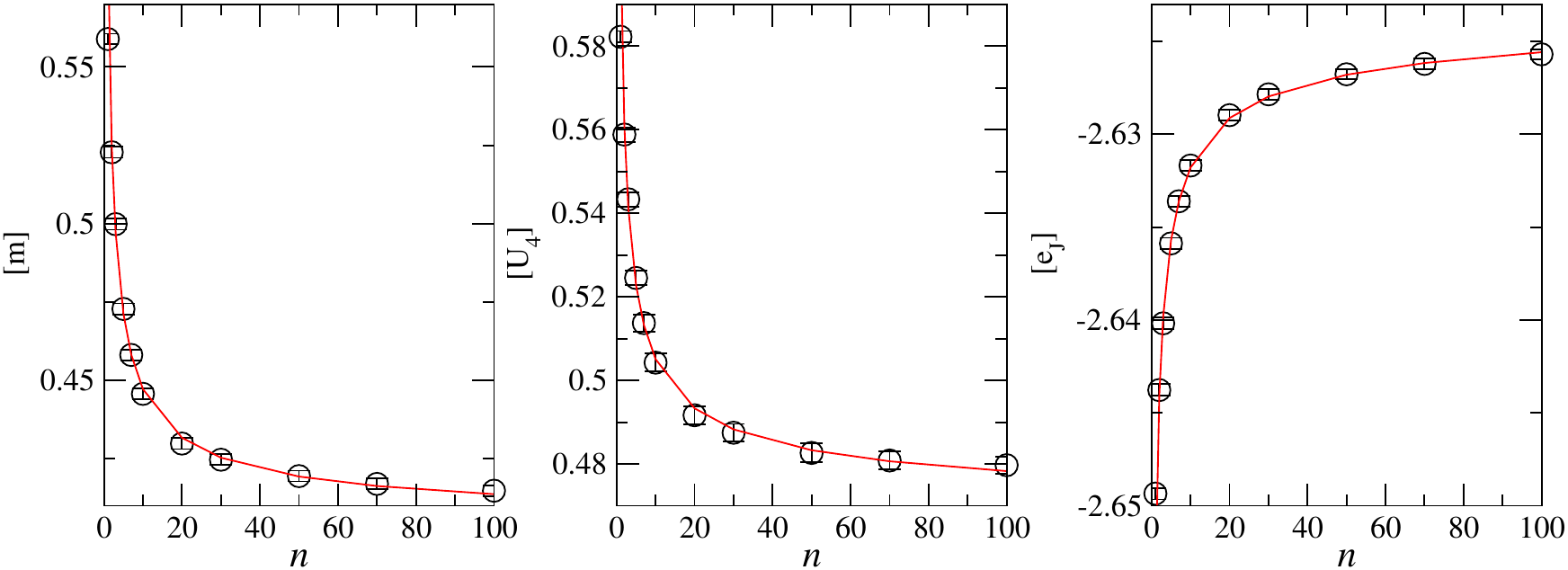}
\caption{Disorder-averaged observables $[m]$, $[U_4]$ and $[e_J]$ as a function of the number of initial conditions $n$ for $\Delta = 1.7$ and $L=64$. The red lines are the simultaneous fits to equations \eqref{extrapolate_m}-\eqref{extrapolate_ej} with $b=0.02$ and $e=0.617\pm0.019$.}
\label{extrapolate}
\end{center}
\end{figure}

We study the behavior of physical quantities as a function of the number of initial conditions $n$. For each disorder realization, we run the simulations for at most 100 different initial spin configurations, i.e., $n_{\max}=100$, and for any $n\le n_{\max}$, we determine all observables from the state of minimum energy out of the $n$ approximate ground states. Fig.~\ref{extrapolate} is a typical plot of disorder-averaged quantities as a function of the number of initial conditions $n$ at fixed $\Delta = 1.7$ and for $L=64$. For the basic observables $m$, $U_4$ and $e_J$ we perform joint fits of the form
\begin{eqnarray}
 m(L,\Delta, n)&=&a_0n^{-b}(1+c_0n^{-e})+m^*(L,\Delta), \label{extrapolate_m}\\
 U_4(L,\Delta, n)&=&a_1n^{-b}(1+c_1n^{-e})+U^*(L,\Delta), \label{extrapolate_U}\\
 -e_J(L,\Delta, n)&=&a_2n^{-b}(1+c_2n^{-e})-e^*(L,\Delta),  \label{extrapolate_ej}
\end{eqnarray}
where $a_0,a_1,a_2,c_0,c_1,c_2$ are individual fit parameters. The power-law exponents $b$ and $e$ are shared between the fits for the three observables. The exponent $b$ is fixed to 0.02. The constants $m^*$, $U^*$, and $e^*$ are the asymptotic values of the magnetization $m$, Binder cumulant $U_4$, and bond-energy per spin $e_J$, respectively. 

\begin{figure}
\begin{center}
\includegraphics[width=0.9\linewidth]{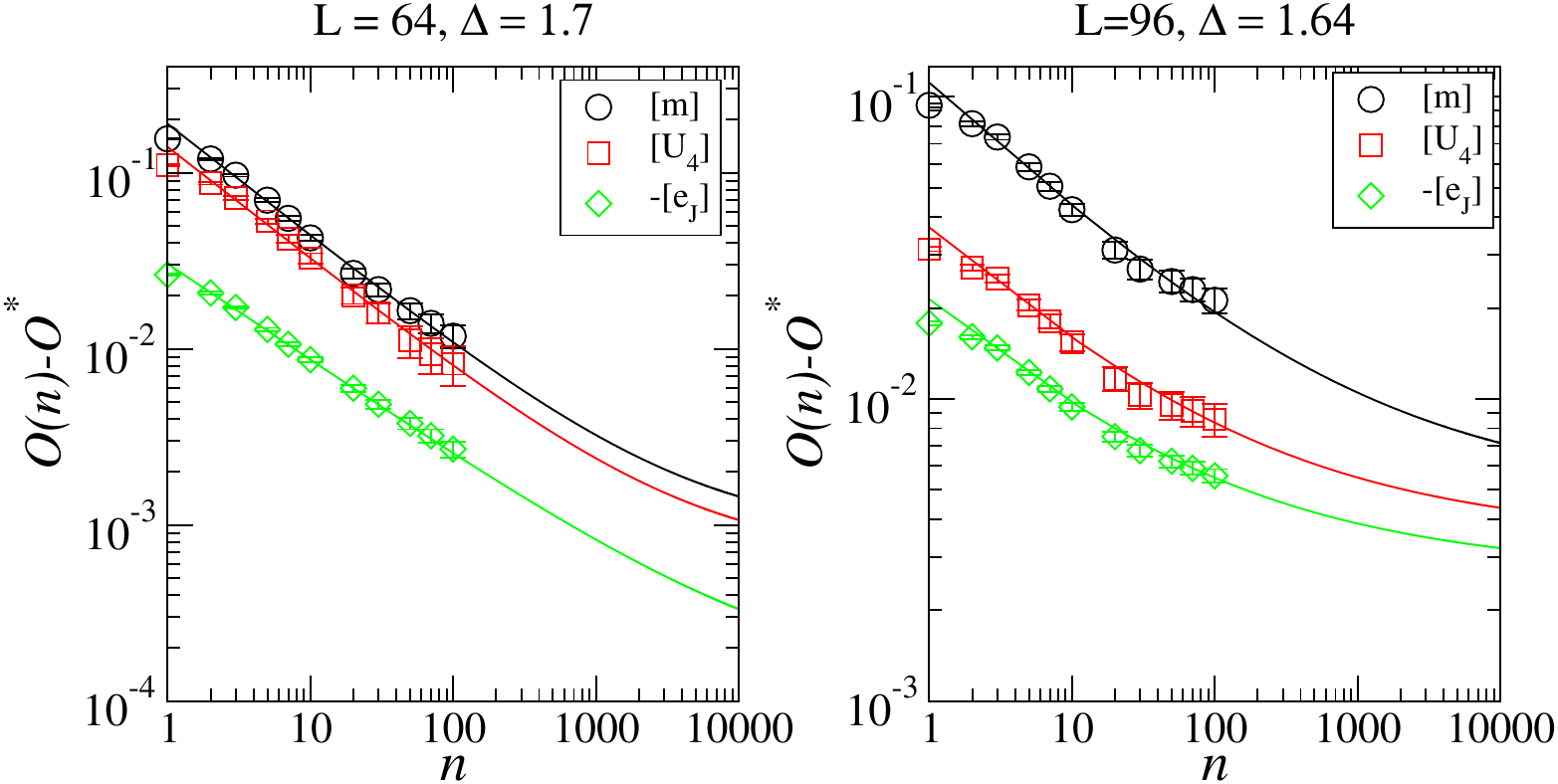}
\caption{ Plot of the residual $\mathcal O(n)-\mathcal O^*$ as a function of $n$ on a double-log scale for  $L=64, \Delta = 1.7$ (left) and $L=96, \Delta = 1.64$ (right). The solid lines are joint fits for the quantities \eqref{extrapolate_m}-\eqref{extrapolate_ej} of the form $an^{-b}(1+cn^{-e})$ with fixed $b=0.02$. The joint fits produce $e=0.617\pm 0.019$ with fit quality $Q=0.73$ for the left panel and $e=0.426\pm 0.028$ with fit quality $Q=0.13$  for the right panel.}
\label{residual}
\end{center}
\end{figure}

\begin{figure}
\begin{center}
\includegraphics[width=0.8\linewidth]{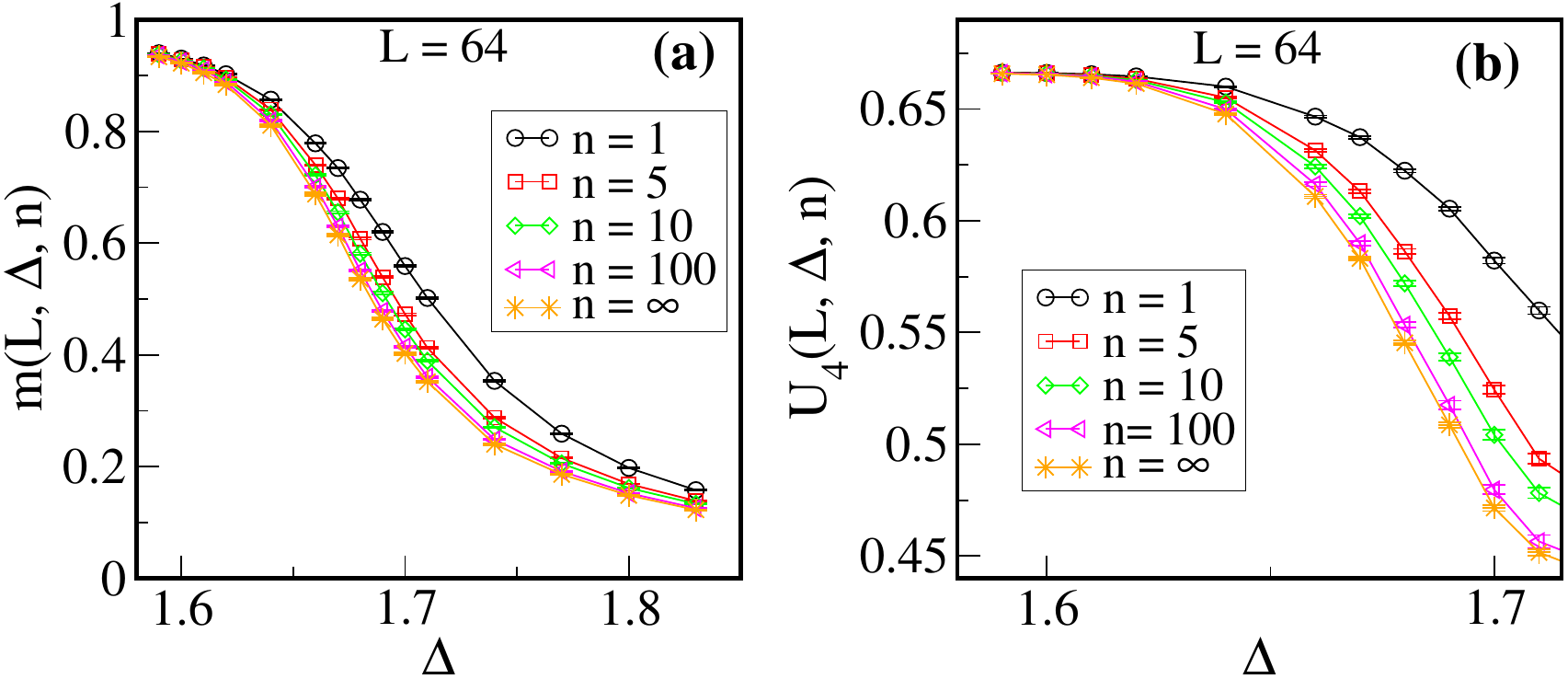}
\caption{ Plots of (a) magnetization $m$ (b) Binder-cumulant $U_4$ averaged over disorder as a function of $\Delta$ for system size $L=64$ and for different number of initial conditions $n$.} 
\label{m_U4}
\end{center}
\end{figure}

Figure~\ref{residual} shows the residuals $\mathcal O(n)-\mathcal O^*$ as a function of $n$, where  the solid lines are the power-law fits. This clearly establishes the fact that these quantities converge to their limiting values in a power-law fashion. In Fig.~\ref{m_U4} we show plots for the disordered averaged $m(L,\Delta, n)$ and $U_4(L,\Delta, n)$ as a function of $\Delta$ but for different $n$. These plots show how the behavior of $m$ or $U_4$ with respect to $\Delta$ modifies as the number of initial conditions $n$ is increased. 

\begin{figure}
\begin{center}
\includegraphics[width=0.9\linewidth]{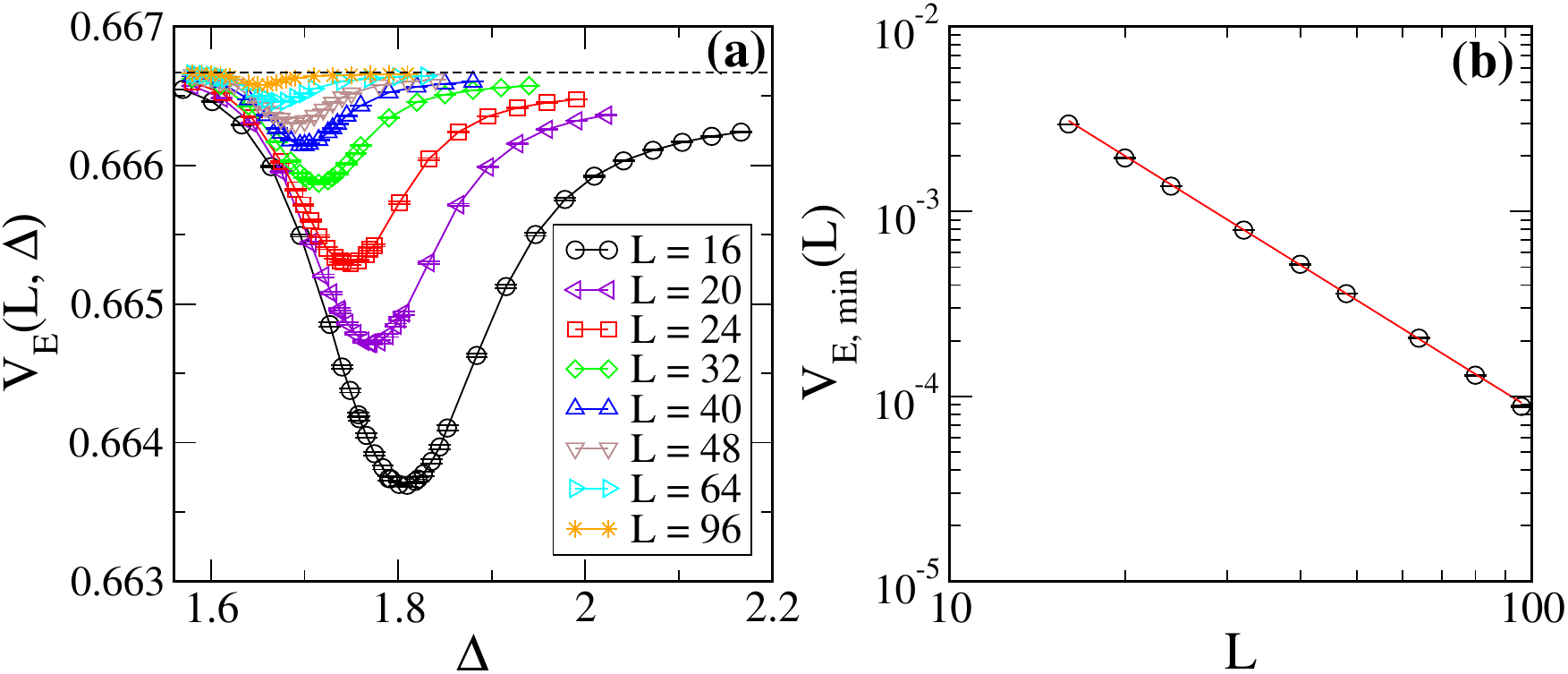}
\caption{ (a) Plots of the energy cumulant $V_E$ against $\Delta$ for various system sizes $L$ (see legend). The dashed line  corresponds to the limiting value 2/3 of $V_E$. (b) Plot of $V_{E,\min}(L)=2/3-V_E(L,\Delta=\Delta_{\min})$ versus $L$ on a log-log scale. The solid line denotes a power-law fit leading to $V_{E,\min}(L)\sim L^{-1.958(8)}$.} 
\label{V_E}<s
\end{center}
\end{figure}

\section{Order of the transition}

As was first pointed out by Binder and co-workers \cite{challa:86,binder:book1}, a useful quantity to study the order of a temperature-driven transition is the energetic cumulant $V_E$ defined as
\begin{equation}
\label{ene_cum}
V_{E} (L,\Delta) = 1 - \frac{[e_J^{4} ]_{\rm av}}{3[e_J^{2}]_{\rm av}^{2}}.
\end{equation}
Panel (a) of Fig.~\ref{V_E} shows a plot of $V_E$ against $\Delta$ for $n=100$ with various system sizes in the range $16\le L\le96$. It can be seen that this quantity has a minimum at a certain $\Delta=\Delta_\mathrm{min}(L)$ and the depth of the minima reduces with increasing $L$. In panel (b), we analyze the minimum of $V_E$, subtracting the trivial limit 2/3, i.e,
\begin{equation}
\label{cum_min}
V_{E,\min} (L) \equiv \frac{2}{3}  - V_{E} (L,\Delta=\Delta_{\min}) .
\end{equation}
We see that the behavior of $V_{E,\min}$ against $L$ follows a power-law decay with the exponent value $1.958\pm0.008$, i.e., $V_{E,\min}(L)\sim L^{-1.958(8)}$.  This suggests a second-order phase transition for the three-state RFPM studied here as $V_{E,\min} \to 0$ in the thermodynamic limit $L\to \infty$  and $V_{E} (\Delta=\Delta_{\min})\to 2/3$. For a first-order transition, $V_E(L,\Delta)$ would be expected to saturate  at a value less than 2/3 \cite{binder:book1}.

\begin{figure}[tb]
\begin{center}
\includegraphics[width=0.98\columnwidth]{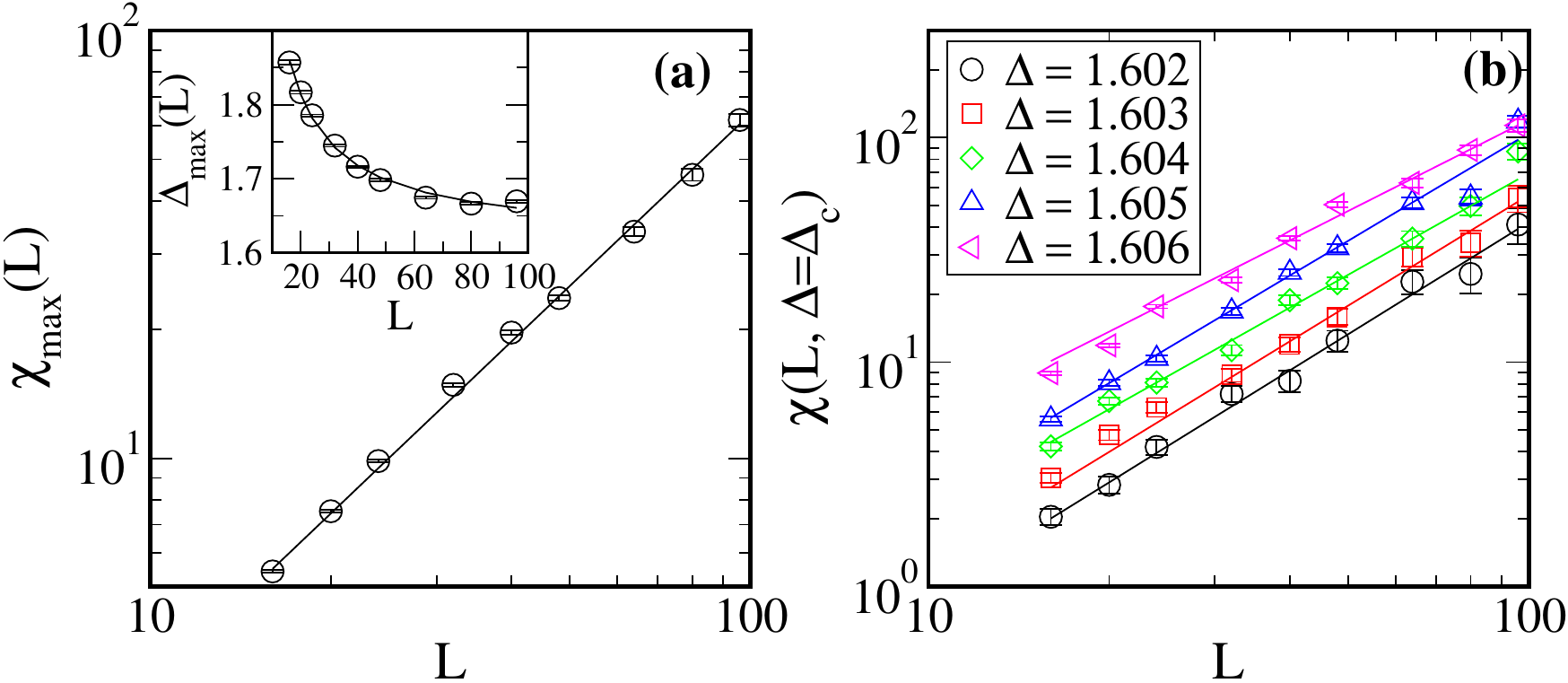}
\caption{(a) Susceptibility maxima $\chi_\mathrm{max}(L)$ for $n=100$ initial conditions according to Eq. (8) of the main paper as a function $L$. The solid line is a power-law fit $\chi_{\max}(L)=AL^{\gamma/\nu}$, yielding $\gamma/\nu = 1.36 \pm 0.01$ with fit-quality $Q=0.034$. The inset shows the location of maxima, i.e., $\Delta_{\max,\chi}(L)$ against $L$. The solid curve corresponds to a fit $\Delta_{\max,\chi}(L) =\Delta_c+a_1 L^{-1/\nu}$ with $\Delta_c=1.621(5)$, $1/\nu=0.97(5)$ and fit-quality $Q=0.007$. (b) $\chi (L, \Delta_c) $  versus $L$ at several $\Delta_c$ for $n=100$. For a better view,
the data for different $\Delta_c$ are shifted relative to each
other by multiplying a constant factor. The data is then fitted to the power-law $\chi (L, \Delta_c)\sim L^{\gamma/\nu}$ for all  $L\ge L_{\min}$ and the results for the exponent $\gamma/\nu$ from different choices of $L_{\min}$ are collected in a Table ~\ref{gamma_exp}. The solid lines show the power-law fits for $L_\mathrm{min} = 16$.}
 \label{chi_deltac}
\end{center}
\end{figure}

\section{Fluctuation formula for the susceptibility}
\label{SI:susceptibility}

We follow the arguments put forward by Schwartz and Soffer \cite{schwartz1985exact} for the
purpose of establishing a bound on the susceptibility exponent of the RFIM to derive
an expression for the zero-field susceptibility of the RFPM. We start from the
Hamiltonian of the model in an additional external magnetic field $H^\alpha$,
\begin{equation}
  \label{hamilt2}
  \mathcal{H}=-J\sum_{\left<ij\right>}\delta_{s_i,s_j}-
  \sum_i\sum_{\alpha=0}^{q-1}(h_{i}^{\alpha}+H^\alpha)\delta_{s_i,\alpha}.
\end{equation}
We consider the magnetization,
\[
  M^\mu = \sum_i \delta_{s_i,\mu},
\]
which in general is a vector of $q$ components.  We are interested in the disorder
average of the derivative $\partial\langle M^\mu\rangle/\partial H^\mu$ in the
limit $H^\mu\to 0$, where $\langle M^\mu\rangle$ denotes the thermal average of
the magnetization.  If we write
\[
  \bar{h}^\alpha_i = h_i^\alpha + H^\alpha,
\]
the disorder average becomes
\begin{equation}
  \left[ \frac{\partial \langle M^\mu\rangle}{\partial H^\mu}\right]_\mathrm{av} =
  \int\d\{\bar{h}_i^\alpha\} P(\{\bar{h}_i^\alpha\})  \frac{\partial \langle
    M^\mu\rangle_{\bar{h}^\alpha_i}}{\partial H^\mu},
  \label{eq:schwartz1}
\end{equation}
where the square brackets indicate averaging over the random fields. Given that
\[
 \frac{\partial \langle M^\mu\rangle}{\partial H^\mu} = \sum_{i,\alpha} \frac{\partial \langle
 M^\mu\rangle}{\partial \bar{h}^\alpha_i} \frac{\partial \bar{h}^\alpha_i}{H^\mu} = \sum_i \frac{\partial \langle
 M^\mu\rangle}{\partial \bar{h}^\mu_i},
\]
partial integration applied to (\ref{eq:schwartz1}) yields
\[
  \frac{\partial \langle M^\mu\rangle}{\partial H^\mu} = -  \sum_i \int\d\{\bar{h}_i^\alpha\}
  \frac{\partial P(\{\bar{h}_i^\alpha\})}{\partial  \bar{h}^\mu_i} \langle M^\mu\rangle_{\{ \bar{h}^\alpha_i\}}.
\]
Since $P(\bar{h}_i^\alpha)$ is a normal distribution of mean $H^\alpha$ and variance
$\Delta$, we have
\[
  \frac{\partial P(\bar{h}_i^\mu)}{\partial \bar{h}_i^\mu} =
  -\frac{\bar{h}_i^\mu-H^\mu}{\Delta^2} P(\bar{h}_i^\mu).
\]
We hence find for the susceptibility,
\begin{equation}
\label{chi}
 \chi^\mu = \lim_{H^\mu\to 0}   \frac{1}{N}\left[ \frac{\partial \langle M^\mu\rangle}{\partial H^\mu}\right]_\mathrm{av} =
  \frac{1}{\Delta^2}\left[\langle m^\mu\rangle \sum_i h_i^\mu\right]_\mathrm{av},
\end{equation}
where $m^\mu = M^\mu/N$ is the magnetization per spin.

While this scheme provides the correct susceptibility in the thermodynamic limit, when spontaneous symmetry breaking occurs, a suitably modified approach is necessary for finite systems. If one interprets $\langle m^\mu\rangle$  literally as magnetization in $\mu$ direction and the corresponding component of the random fields is sampled, there is not explicit symmetry breaking, and the resulting estimator does not show a maximum. A similar result is achieved when using the sample-dependent maximum definition of $\langle m^\mu\rangle$ and the corresponding component of the random fields. An explicit symmetry breaking can be achieved by applying a small external field; asymptotically, the total energy contribution of an external field of magnitude $H$ is $H N$, where $N$ is the number of spins. The energy scale of the contribution of the random fields deep in the ferromagnetic phase is $\Delta\sqrt{N}$. Hence a field of strength $H \propto \Delta/\sqrt{N}$ is sufficient to break the symmetry.

Here, we stress that the susceptibility $\chi$ can also be studied by a standard approach of determining the GS explicitly in the presence of external fields $H$ and fitting $[m]_{\rm av}$ against $H$ to a function $y=a_0+a_1x+a_2x^2$. The parameter $a_1$ then determines  $\chi$. While this works relatively well for the RFIM if exact GS algorithms are used \cite{ahrens:11a}, for the RFPM the approximate nature of the algorithm can lead to negative values of $\chi$ and other artifacts.
 
Applying the explicit symmetry breaking described above and the parabolic fit procedure described in the main paper, we arrive at estimates for the maxima of the susceptibility which are expected to scale as
\begin{equation}
 \chi_{\max}(L) \sim L^{\gamma/\nu}, 
\end{equation}
while the locations of the maxima behave as
\begin{equation}
\Delta_{\max,\chi}(L)=\Delta_c+a_1L^{-1/\nu}.
\end{equation}
Our simulation data presented in Fig.~\ref{chi_deltac}(a) are consistent with these assumptions, but the somewhat large value for $1/\nu$ hints at the presence of unresolved scaling corrections. For the alternatively performed analysis at fixed random-field strengths $\Delta \in [1.602,1.606]$ results are shown in Fig.~\ref{chi_deltac}(b). As the fit parameters in Table \ref{gamma_exp} imply, the dependence on the fixed $\Delta$ used is quite moderate in this range, and we hence use the central value for $\Delta = 1.604$, which is $\gamma/\nu = 1.51(6)$.
 
% Notes:
 %\begin{itemize}
  %   \item We have studied RFIM susceptibility extensively through this definition. The results %will be presented elsewhere.
   %  \item The susceptibility $\chi$ can also be studied by a standard approach of determining the GS explicitly in the presence of external-fields $H$ and fitting $[m]_{\rm av}$ against $H$ to a function $y=a_0+a_1x+a_2x^2$. The parameter $a_1$ then determines  $\chi$. However, this doesn't work for the RFPM due to the approximate GS whereby $[m]_{\rm av}$ at any $H$ can be lower than that at previous $H$, resulting in a negative value of $\chi$. 
 %\end{itemize}

\begin{table*}[tb]
\centering
\begin{ruledtabular}
\begin{tabular}{ c  c  c c c c c c c c c }
 
$L_{\min}$ & \multicolumn{2}{c}{$\Delta_c=1.602$} & \multicolumn{2}{c}{$\Delta_c=1.603$} &\multicolumn{2}{c}{$\Delta_c=1.604$} & \multicolumn{2}{c}{$\Delta_c=1.605$} &\multicolumn{2}{c}{$\Delta_c=1.606$}\\
\cline{2-3} \cline{4-5} \cline{6-7} \cline{8-9} \cline{10-11}
&$\gamma/\nu$&$Q$&$\gamma/\nu$&$Q$&$\gamma/\nu$&$Q$&$\gamma/\nu$&$Q$&$\gamma/\nu$&$Q$\\ \hline
16& 1.64(6)& 0.47 &1.54(6)&0.49&1.512(58)&0.68&1.59(6)&0.67&1.463(61)&0.67\\ 
20&1.635(73)& 0.36 &1.512(74)&0.43&1.467(72)&0.71&1.583(74)&0.56&1.453(76)&0.56
\\ 
24& 1.58(9)& 0.35 &1.60(9)& 0.57&1.493(91)& 0.61&1.595(92)&0.43&1.380(96)&0.66 \\ 
32& 1.56(13)&0.24&1.69(13)& 0.55&1.62(13)& 0.78&1.55(13)& 0.33&1.44(14)&0.58\\
\end{tabular}
\end{ruledtabular}
    \caption{The exponent $\gamma/\nu$ from fits of the form $\chi(L,\Delta_c) \sim L^{\gamma/\nu}$ for different estimates $\Delta_c$ of the critical field strength as a function of the cut-off $L_{\min}$ (data for $n=100$).}
 \label{gamma_exp}
\end{table*}

\bibliography{ref.bib}